\documentclass[twocolumn,superscriptaddress,showpacs,preprintnumbers,amsmath,amssymb,prb]{revtex4-1}
\usepackage{verbatim}
\usepackage{epsfig}
\usepackage{graphicx}

\usepackage{color}
\usepackage[colorlinks,bookmarks=false,citecolor=blue,linkcolor=red,urlcolor=blue]{hyperref}
\usepackage{dcolumn}
\usepackage{bm}

\begin{document}

\title{Impurity induced phase competition and supersolidity}

\author{Madhuparna Karmakar}
\email{madhuparna.k@gmail.com}
\affiliation{The Institute of Mathematical Sciences, HBNI, C I T Campus, Chennai 600 113, India}
\affiliation{Indian statistical institute, Chennai centre, MGR knowledge city,\\
CIT campus, Taramani, Chennai 600113, India}
\affiliation{Department of Physics, Indian Institute of technology, Madras, Chennai 600036, India.}
\author{R. Ganesh}
\email{ganesh@imsc.res.in}
\affiliation{The Institute of Mathematical Sciences, HBNI, C I T Campus, Chennai 600 113, India}
\date{\today}

\begin{abstract}
Several material families show competition between superconductivity and other orders. When such competition is driven by doping, it invariably involves spatial inhomogeneities which can seed competing orders. 
We study impurity-induced charge order in the attractive Hubbard model, a prototypical model for competition between superconductivity and charge density wave order.
We show that a single impurity induces a charge-ordered texture over a length scale set by the energy cost of the competing phase. 
Our results are consistent with a strong-coupling field theory proposed earlier in which 
superconducting and charge order parameters form components of an $SO(3)$ vector field. 
To discuss the effects of multiple impurities, we focus on two cases: correlated and random distributions. In the correlated case, the CDW puddles around each impurity overlap coherently leading to a `supersolid' phase with coexisting pairing and charge order. In contrast, a random distribution of impurities does not lead to coherent CDW formation. 
We argue that the energy lowering from coherent ordering can have a feedback effect, driving correlations between impurities. This can be understood as arising from an RKKY-like interaction, mediated by impurity textures. 
We discuss implications for charge order in the cuprates and doped CDW materials such as NbSe$_2$. 
\end{abstract}
\maketitle
\section{Introduction:}
Many unconventional superconductors arise from a competing host phase when it is destabilized by doping. Upon tuning the dopant concentration, the competing phase progressively weakens and gives way to a superconducting dome. This is seen in the phase diagram of the cuprates (with competing N\'eel order)\cite{cupratereview}, pnictides (stripe order)\cite{pnictidecoexistence}, TiSe$_2$ (charge density wave order)\cite{Morosan2006}, etc.
In addition, a recent wave of experiments on the cuprates has uncovered charge order within the superconducting dome, which is highly sensitive to doping. In all these systems, at first glance, the effect of doping is simply to change the number of carriers in the system. However, doping invariably involves impurity atoms that form a disordered background. Can this intrinsic disorder lead to observable macroscopic consequences? We discuss this question, with a view to using disorder as a tool to manifest competing orders. We place our discussion in the context of the attractive Hubbard model, the simplest model to show competition between superconductivity (SC) and charge density wave (CDW) orders. 
\subsection{Hubbard model:}
We consider fermions on a square lattice with an on-site attractive interaction, with the Hamiltonian
\begin{equation}
 H_{t,U} \!=\! -t \sum_{\langle ij \rangle,\sigma} \left\{ c_{i,\sigma}^\dagger c_{j,\sigma} +
 h.c.\right\} -U\sum_{i}\hat{n}_{i,\uparrow} \hat{n}_{i,\downarrow} - \mu \sum_{i,\sigma} \hat{n}_{i,\sigma}
 \label{eq.Hamiltonian}
 \end{equation}
This model possesses a remarkable $SO(3)$ symmetry when $\mu$ is tuned to half-filling and the hopping is restricted to nearest-neighbors. At this fine-tuned point, SC becomes degenerate with CDW order\cite{Yang1989,Zhang1990,Yang1990,Burkov2008,Ganesh2009}. These two orders combine to form an enlarged order parameter space with $SO(3)$ character\cite{Burkov2008,Ganesh2009,Karmakar2017}. In this study, we break this degeneracy by introducing a next-nearest neighbor hopping $t'$. As a result, we are left with a SC ground state and low lying CDW excitations. The $SO(3)$ character is clearly revealed at strong coupling when a pseudospin description arises. The model maps to the Heisenberg antiferromagnetic Hamiltonian, with small corrections arising from $t'$ terms\cite{Burkov2008,Ganeshthesis}. 
\begin{figure}
\includegraphics[width=\columnwidth]{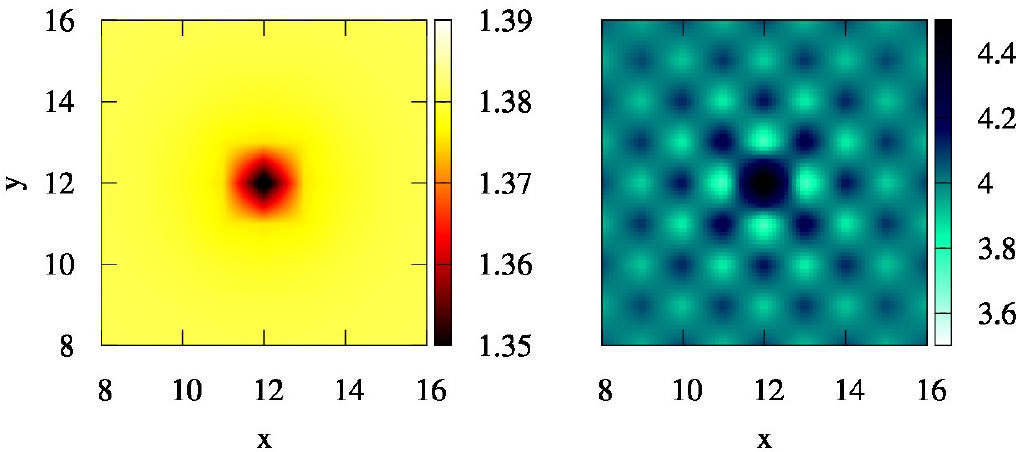}
 \caption{Texture induced by a single impurity: spatial maps of SC (left) and density (right) order parameters. The data corresponds to $U = 4t$, $t'=0.1t$ and $W = 0.1t$. }
\label{fig.single_map}
\end{figure}
\subsection{Strong coupling field theory}
In an earlier work, we proposed that the Hubbard Hamiltonian at strong coupling maps on to a strongly-coupled field theory with a constant-squared-sum constraint\cite{Karmakar2017}. Here, we provide further evidence for this mapping by evaluating impurity-induced texture using both approaches. The two approaches come out to be in excellent agreement.
The field theory (the Landau Ginzburg free energy density) in terms of the SC and CDW order parameters, $\Delta(\mathbf{r})$ and $\tilde{\phi}(\mathbf{r})$, is given by
\begin{eqnarray}
\nonumber \mathcal{L} = \frac{\rho}{2} \left| \left( \mathbf{\nabla} - \frac{ie}{\hbar c}\mathbf{A} \right) \Delta(\mathbf{r} ) \right|^2
+ \frac{1}{8\pi} \left( \mathbf{\nabla} \times \mathbf{A} \right)^2 \\
+ \frac{\rho}{2} \vert \mathbf{\nabla}\tilde{\phi}(\mathbf{r} ) \vert^2
-\chi \vert \Delta (\mathbf{r} )\vert^2 - \chi(1- gt'^2)\vert \tilde{\phi} (\mathbf{r} )\vert^2.
\label{eq.LG}
\end{eqnarray}
The order parameters are not to be taken as independent. They are strongly coupled by a constraint 
\begin{equation}
\vert \Delta (\mathbf{r} ) \vert^2 + \tilde{\phi}^2(\mathbf{r} ) = c^2,
\label{eq.constraint}
\end{equation}
a constant. This is analogous to the hypothesized $SO(5)$ model for the cuprates which combines SC and antiferromagnetism into a five-dimensional order parameter vector. The enlarged order parameter vector is taken to have a constant amplitude independent of time and space\cite{Arovas1997,Demler2004}. More recently, several similar theories have been proposed to explain charge ordering in the cuprates\cite{Efetov2013,Hayward2014,Wachtel2014}. Our work demonstrates a microscopic origin for such a constant-squared-sum constraint. While we focus on the Hubbard model, our results are broadly applicable to other systems described by this field theory. 

As can be seen from Eq.~\ref{eq.LG}, the $t'$ coupling breaks the $SO(3)$ symmetry. It leads to a SC ground state with low lying CDW fluctuations. 
In this context, we study the CDW texture induced by impurities. We focus on the strong coupling limit of the Hubbard model where the field theory in Eq.~\ref{eq.LG} is expected to hold. 

\section{Single impurity response}
\label{sec.single_imp}

\begin{figure}
\includegraphics[width=\columnwidth]{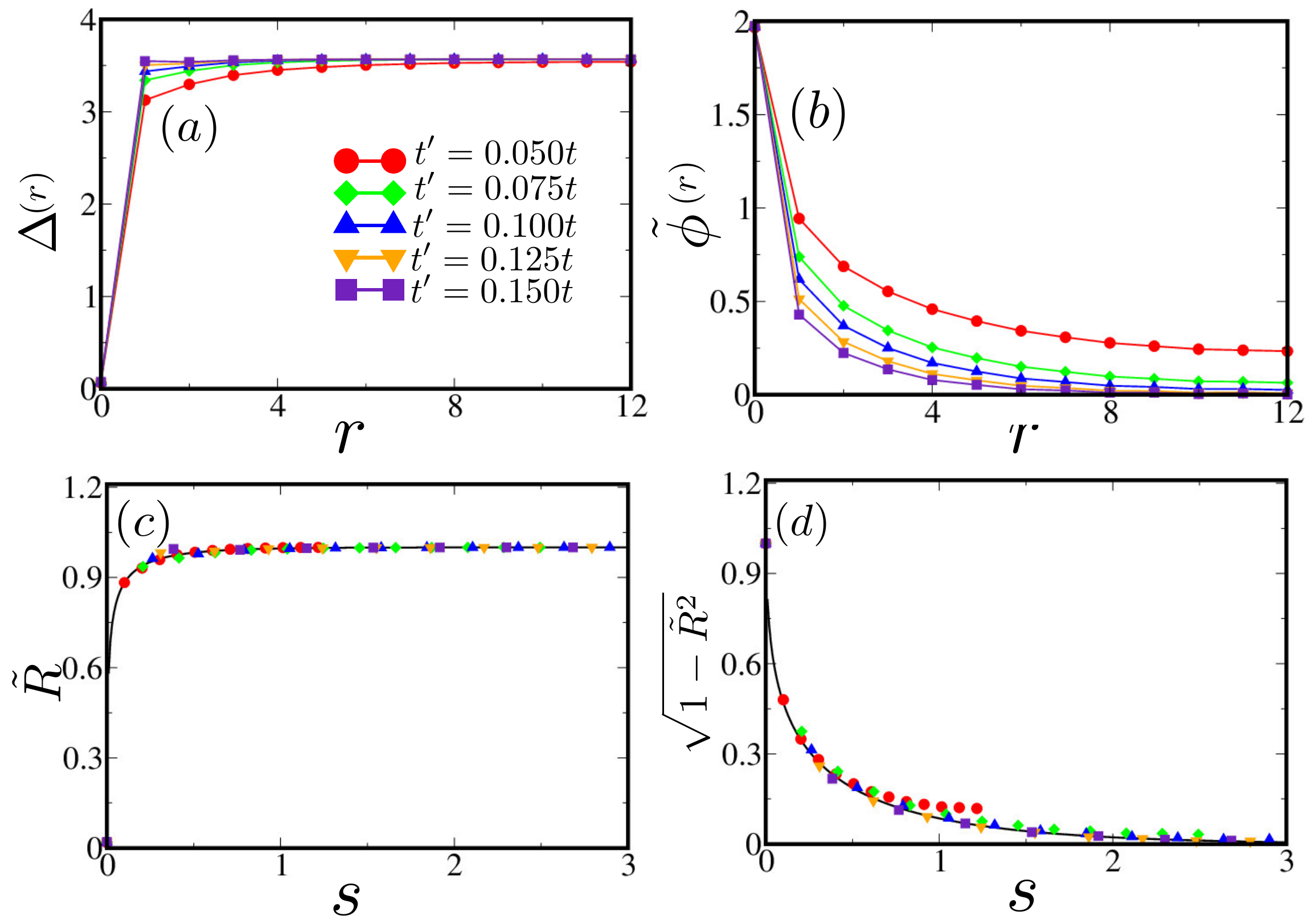}
 \caption{Texture induced by a single impurity: SC (top left) and CDW (top right) order parameter profiles. The data is extracted from mean field simulations with $U = 4t$, $W = 10t$ and various $t'$ values. The bottom two panels show the same data after rescaling (see text). The data for different $t'$ collapse onto the black line, which is the result obtained from the strong coupling field theory.
 }
\label{fig.single_imp_profiles}
\end{figure}
We had previously studied the role of a magnetic field 
in the Hubbard model\cite{Karmakar2017}. We showed that vortex cores contain puddles of CDW order. At a critical field strength, vortex cores overlap to give rise to a coexistence phase -- a `supersolid'. In this study, we consider the role of impurities in bringing out this phase competition. Specifically, we focus on on-site impurity potentials.
When a tendency towards CDW order exists, it is immediately clear that an on-site potential will locally induce this order.
Depending on the sign of the potential, it favors larger or smaller density on the impurity site. This can also be seen in the strong coupling pseudospin model: an on-site potential maps to a local field that couples to the $z$-moment of the $SO(3)$ pseudospin\cite{Burkov2008,Ganeshthesis}. 
Here, we first study the CDW texture induced by single on-site impurity.
We then discuss how impurity-induced CDW correlations can overlap, leading to phase coexistence. 
\begin{figure*}
\includegraphics[width=3.5in]{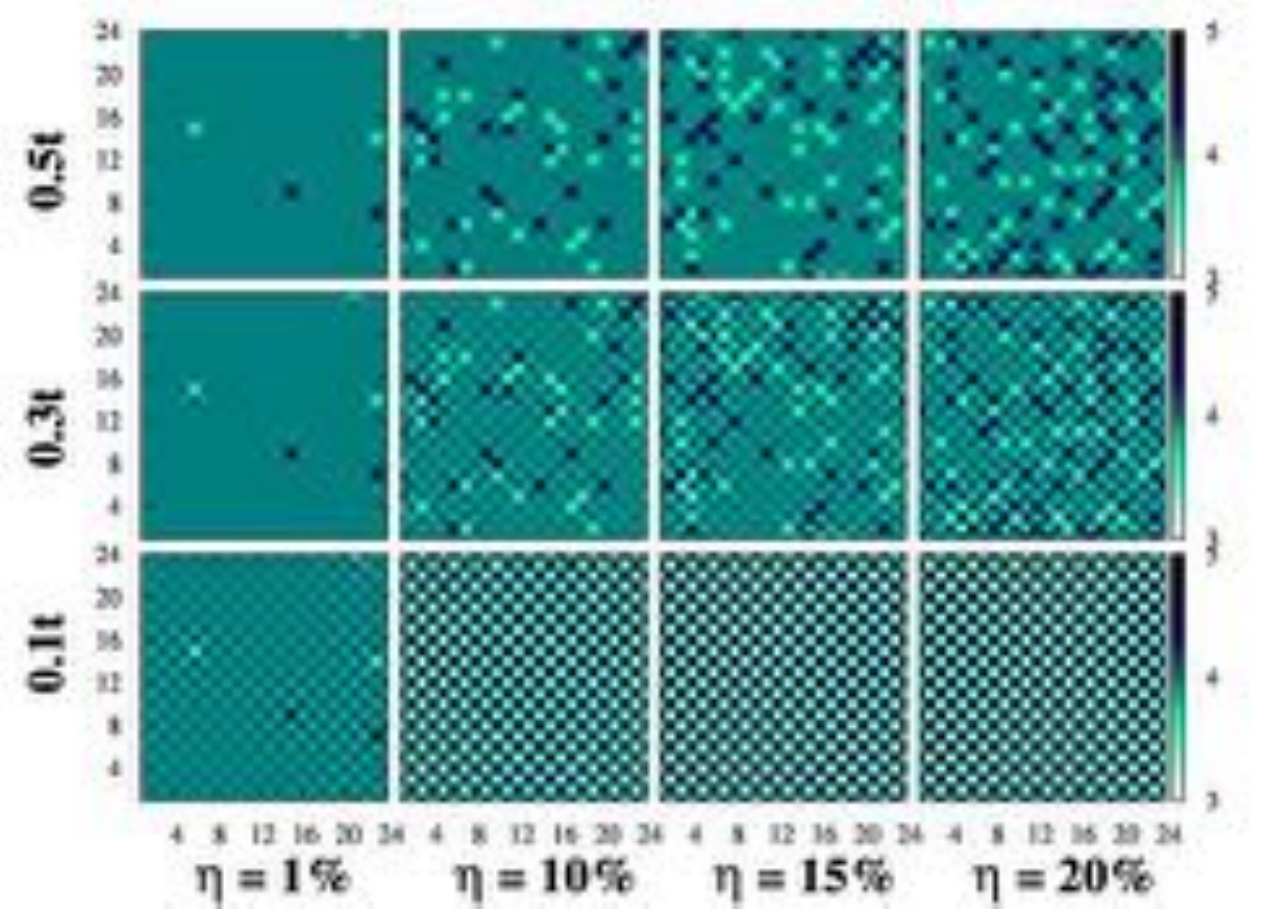}
\includegraphics[width=3.5in]{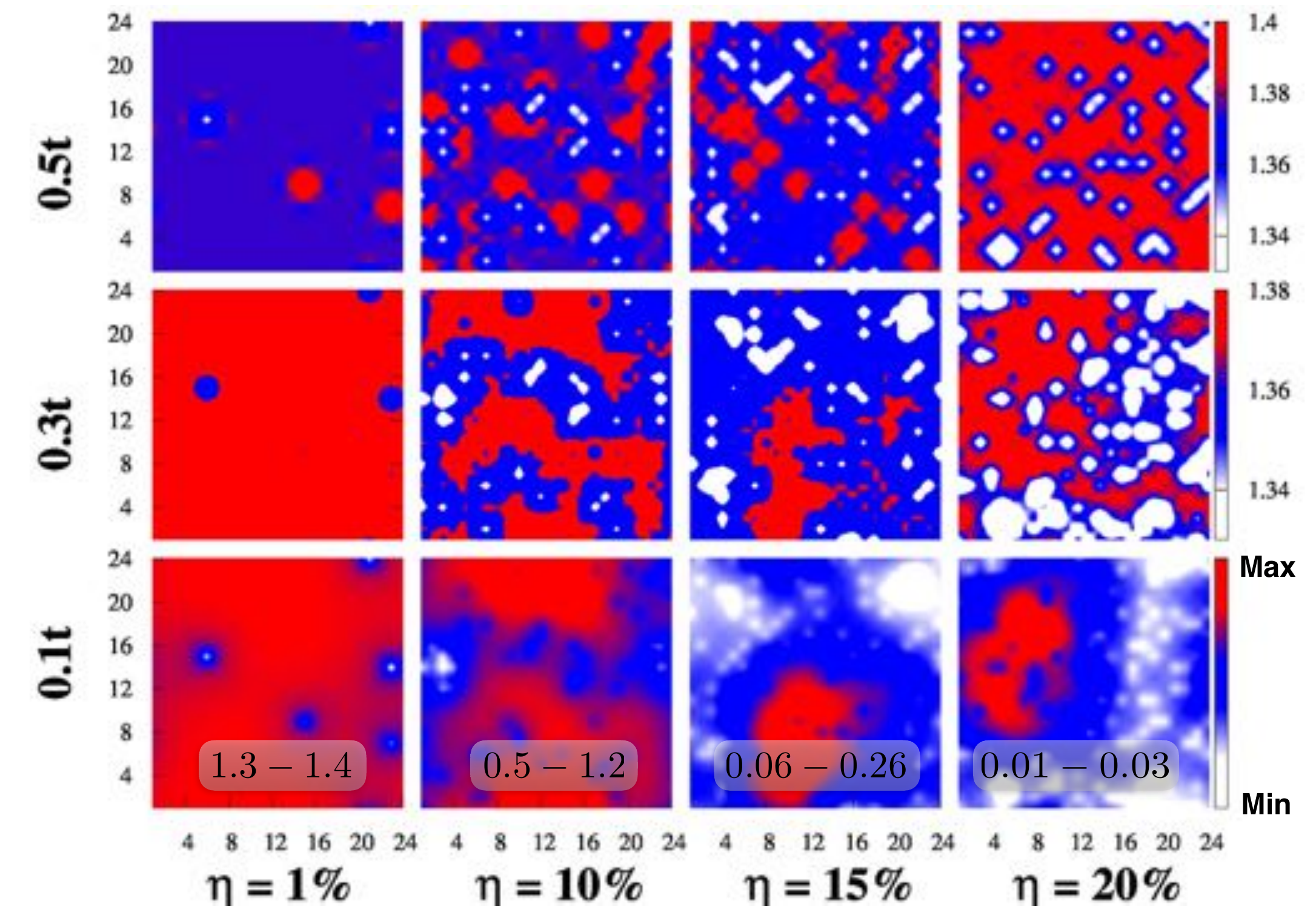} 
 \caption{Evolution of order parameters with $t'$ and impurity concentration, $\eta$, with correlated impurities: (Left) Maps of density. The rows and columns correspond to fixed $t'$ and $\eta$ values respectively. (Right) Maps of $\Delta$, the SC order parameter. 
All panels show data corresponding to $U=4t$ and $W=0.1t$ on a $24\times 24$ lattice. Each plot shows the mean-field results obtained from a sample disorder configuration.  
In the bottom row of panels on the right, the color bar range is mentioned in each panel separately.}
\label{fig.corr_maps}
\end{figure*}

\subsection{Texture from field theory}
We first examine the effect of a single impurity in the field theory given by Eq.~\ref{eq.LG}. Our analysis is similar to that of Ref.~\onlinecite{Arovas1997} which discusses vortex structure within the $SO(5)$ model of the cuprates.  
We assume a strong local potential which induces maximal CDW order at the impurity site. Correspondingly, the SC amplitude will be suppressed to zero, in keeping with the constraint of Eq.~\ref{eq.constraint}. As we move away from the impurity, we expect the CDW amplitude to decay and the SC amplitude to recover to its uniform value. 
In the free energy expression, we set $\mathbf{A}=0$ as we do not have a magnetic field. Further, we set $\Delta(\mathbf{r}) = R(\mathbf{r})$, a real-valued field, as we do not expect the SC phase to vary. Our impurity problem has rotational symmetry at length scales greater than the lattice spacing; we further assume that $R(\mathbf{r})$ depends only on the radial coordinate $r$, with the impurity located at the origin.  
Using the constraint described above, we now rewrite the CDW order parameter as $\tilde{\phi}(r) = \sqrt{c^2-R^2(r)}$.
We obtain
\begin{eqnarray}
\mathcal{L} \sim \frac{\rho}{2} \left| \mathbf{\nabla}  R(r ) \right|^2
+ \frac{\rho}{2} \vert \mathbf{\nabla}\sqrt{c^2-R^2(r)} \vert^2
- \chi gt'^2 R^2(r ).\phantom{ab}
\label{eq.LG1}
\end{eqnarray}
This action contains two scales: $c$, an order-parameter scale, and $\xi =  \sqrt{\frac{\rho}{2\chi gt'^2}}$, a length scale. To make the action dimensionless, we redefine $\tilde{R}(\mathbf{r}) = R(\mathbf{r})/c$ and rescale the position coordinate using $\mathbf{s}=\mathbf{r}/\xi$. The saddle point equation for the resulting action is
\begin{eqnarray}
\frac{d^2 \tilde{R}}{ds^2} + \frac{1}{s}\frac{d \tilde{R}}{d s} 
+\frac{\tilde{R}}{1-\tilde{R}^2}\left(\frac{d \tilde{R}}{d s}\right)^2 + (1-\tilde{R}^2)\tilde{R} = 0.\phantom{acb}
\label{eq.LG2}
\end{eqnarray}
As this differential equation does not have an analytic solution, we find solutions using the shooting method. We assume that the impurity potential completely suppresses $\Delta(\mathbf{r})$ at the origin, providing a boundary condition, $\tilde{R}_{s=0}=0$. The second boundary condition is given by $\tilde{R}_{s\rightarrow \infty }=1$, as we have uniform SC order far away from the impurity. Imposing $\tilde{R}_{s=0}=0$, we guess a suitable value for $\tilde{R}'_{s=0}$ which satisfies the other boundary condition. The obtained curve for $\tilde{R}(s)$ is plotted as the black line in Fig.~\ref{fig.single_imp_profiles} (bottom left).

\begin{figure*}
\includegraphics[width=2\columnwidth]{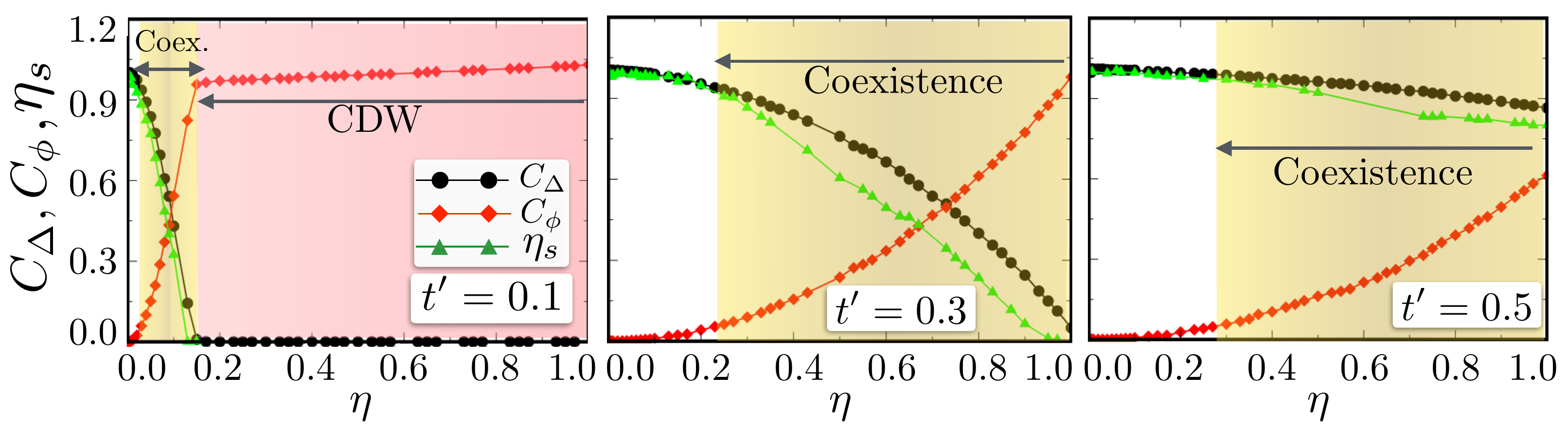}
 \caption{Superfluid stiffness ($\eta_{s}$), SC and CDW correlation functions as a function of impurity concentration, $\eta$, for three $t'$ values. We have not shown error bars for clarity; all error bars are less than $0.2$. }
\label{fig.Cdelta_phi}
\end{figure*}

\subsection{Mean field results}
To compare with the field theory result, we study the impurity texture using Bogoliubov deGennes (BdG) mean field theory. We perform a quadratic decomposition of the Hubbard interaction in both pairing and density channels\cite{Karmakar2017} with order parameters $\Delta_i = U \langle c_{i\uparrow} c_{i\downarrow} \rangle$ and $\phi_i = \langle c_{i\uparrow}^\dagger c_{i\uparrow} + c_{i\downarrow}^\dagger c_{i\downarrow}\rangle$. Working on an $L\times L$ lattice with periodic boundaries, we evaluate the order parameters self-consistently. 
We restrict ourselves to half-filling and strong coupling ($U \gg t,t'$), the regime where the field theory of Eq.~\ref{eq.LG} is expected to apply.
Indeed, our mean-field results are in agreement with squared-sum constraint. For $U = 4t$, we find $\vert \Delta (\mathbf{r}) \vert^2 + \vert \tilde{\phi}(\mathbf{r})\vert^2$ to be constant within a $3\%$ error margin. For $U=8t$, the corresponding error margin is $1\%$. 

We find that a single impurity indeed induces a CDW texture, as shown in Fig.~\ref{fig.single_map}. The SC and CDW profiles around a single impurity are shown for different values of $t'$ in the upper panels of Fig.~\ref{fig.single_imp_profiles}. 
To compare with the field theory result, we scale the order parameters by $c=\Delta_0$, the amplitude of the SC order parameter in the ground state of the impurity-free problem. 
We also rescale position by $\xi =  \sqrt{\frac{\rho}{2\chi gt'^2}}$. We determine the parameters $\rho/2$ and $\chi g t'^2$ from our mean field simulations as follows.

To determine $\rho/2$, we induce a flowing-SC mean-field state with $\Delta(\mathbf{R}) = \Delta_0 e^{i\mathbf{Q}\cdot\mathbf{r}}$. The wavevector of flow, $\mathbf{Q}$, is chosen to be small so that the phase of the SC order parameter winds slowly. 
The increase in energy due to flow is $\Delta E_{flow} = \frac{\rho}{2} \vert \mathbf{Q} \vert^2 \Delta_0^2$. This allows us to extract $\rho/2$. To find $\chi gt'^2$, we compare the energies of the uniform SC state and the uniform CDW state. The difference in energies is precisely $\chi g t'^2$. This allows us to extract $\xi$ from our mean field theory for any given values of the microscopic parameters $U$ and $t'$. Note that $\xi \sim 1/t'$, i.e., the size of the impurity texture scales inversely with the $SO(3)$-breaking term.


Fig.~\ref{fig.single_imp_profiles} shows our mean-field results for the CDW texture around single impurity. After scaling the order parameters and distance appropriately, we find that the data for various $t'$ values collapses onto a single line. Moreover, this line is precisely the field theory result obtained by the shooting method. This constitutes strong evidence that the Hubbard model indeed maps onto the field theory given in Eq.~\ref{eq.LG}.

\section{Multiple correlated impurities}
\label{sec.correlated}

Having established that impurities seed local puddles of CDW order, we next focus on multiple impurities. 
In this section, we focus on `correlated' impurities -- we take the Hamiltonian to be of the form,
\begin{eqnarray}
H_{corr.} =H_{t,t',U} + W \sum_{j_r (\eta)}(-1)^{j_r} c_{j_r,\sigma}^\dagger c_{j_r,\sigma}. 
\end{eqnarray}
Here, $W$ is a fixed positive quantity representing the strength of each impurity potential. 
The index $j_r$ sums over a randomly chosen set of sites, with all quantities averaged over 100 disorder realizations. We treat the the impurity concentration, $\eta$, as a tuning parameter. 
The factor $(-1)^{j_r}$ forces the induced CDW to be of the same kind at every impurity location -- this ensures that CDW textures can overlap coherently. In Sec.~\ref{sec.corr_egy_gain} below, we provide an \textit{a posteriori} justification for choosing this impurity scheme. We perform a Bogoliubov deGennes mean-field analysis at half-filling, once again decoupling the interaction in pairing and density channels.

Figure.~\ref{fig.corr_maps} shows the spatial maps of the pairing and density order parameters. Each map is for a sample disorder configuration for given values of $t'$ and $\eta$. It is evident that each impurity induces local CDW correlations. As expected, the size of each CDW puddle decreases with increasing $t'$. This is consistent with the field theory analysis which reveals that the impurity texture has a length scale that is inversely proportional to $t'$. 
As the disorder concentration, $\eta$, is increased, the puddles overlap, giving rise to long ranged CDW correlations which coexist with SC order. In addition, the SC order parameter progressively diminishes with increasing impurity concentration. 

To quantify the changes in the order parameters, we define two quantities $C_\Delta = \frac{1}{\Delta_0^2 L^2}\sum_{i,j}\Delta_i \Delta_j^*$ and $C_\phi = \frac{2}{L^2 }\sum_{i,j}\tilde{\phi}_i \tilde{\phi}_j$. Here, $\Delta_0$ is the SC amplitude obtained in the clean limit. $\tilde{\phi}_i = (-1)^i(\phi_i - \bar{\phi})$ is the local CDW order parameter, with $\bar{\phi}$ being the average density which is unity at half-filling. $C_\Delta$ is designed to go to unity in the clean limit. 
Fig.~\ref{fig.Cdelta_phi} plots these correlation functions vs. impurity concentration for different values of $t'$. As expected, $C_\phi$ increases with $\eta$, while $C_{\Delta}$ decreases. 

\subsection{Phase coexistence}
As impurity textures overlap, CDW correlations extend over longer length scales. To identify the onset of long range CDW order, we introduce a heuristic criterion $C_{\phi}\gtrsim 0.05$. In our mean-field simulations, we find that this criterion corresponds to the point at which CDW correlations first contiguously stretch over the entire system. 
In Fig.~\ref{fig.Cdelta_phi}, we have used this criterion to mark the onset of the `Coexistence' region. In addition, this region shows long ranged SC order as indicated by the non-zero values of $C_{\Delta}$. 

The coexistence of long ranged density order and pairing constitutes `supersolidity'\cite{Boninsegni2012}. We had previously demonstrated supersolidity in the clean limit of the Hubbard model, but in the presence of a magnetic field. The magnetic field creates vortices with charge ordered cores. The overlap of vortices give rise to phase coexistence. Here, an analogous role is played by on-site impurities. A crucial difference is that a vortex does not have a preference for one of the two kinds of checkerboard CDW order. However, with an impurity, the sign of the potential determines the nature of the induced CDW order. This leads to marked differences between correlated and random disorder schemes. With correlated impurities, we do find phase coexistence. 

To check for the stability of the supersolid phase, we measure superfluid stiffness as the energy cost incurred due to a gradual winding of the SC phase\cite{castellani2012}. We introduce a vector potential $\mathbf{A} = 2\pi/L \hat{x}$, which does not lead to a magnetic field piercing the lattice. Rather, it induces a phase winding of $4\pi$ from one end of the cluster to the other, i.e., $\Delta{\mathbf{r}}\sim \Delta_0 e^{i\mathbf{Q}\cdot \mathbf{r}}$ with $\mathbf{Q}=4\pi \hat{x}/L $. The resulting increase in energy, averaged over disorder realizations, is identified as the superfluid stiffness. The values of stiffness, obtained using a $24\times 24$ lattice, are plotted in Fig.~\ref{fig.Cdelta_phi}. We find that stiffness approximately tracks $C_\Delta$ for all $t'$ and $\eta$ values. The stiffness is non-zero within the coexistence phase, indicating stability against fluctuations.

\begin{figure}
\includegraphics[width=\columnwidth]{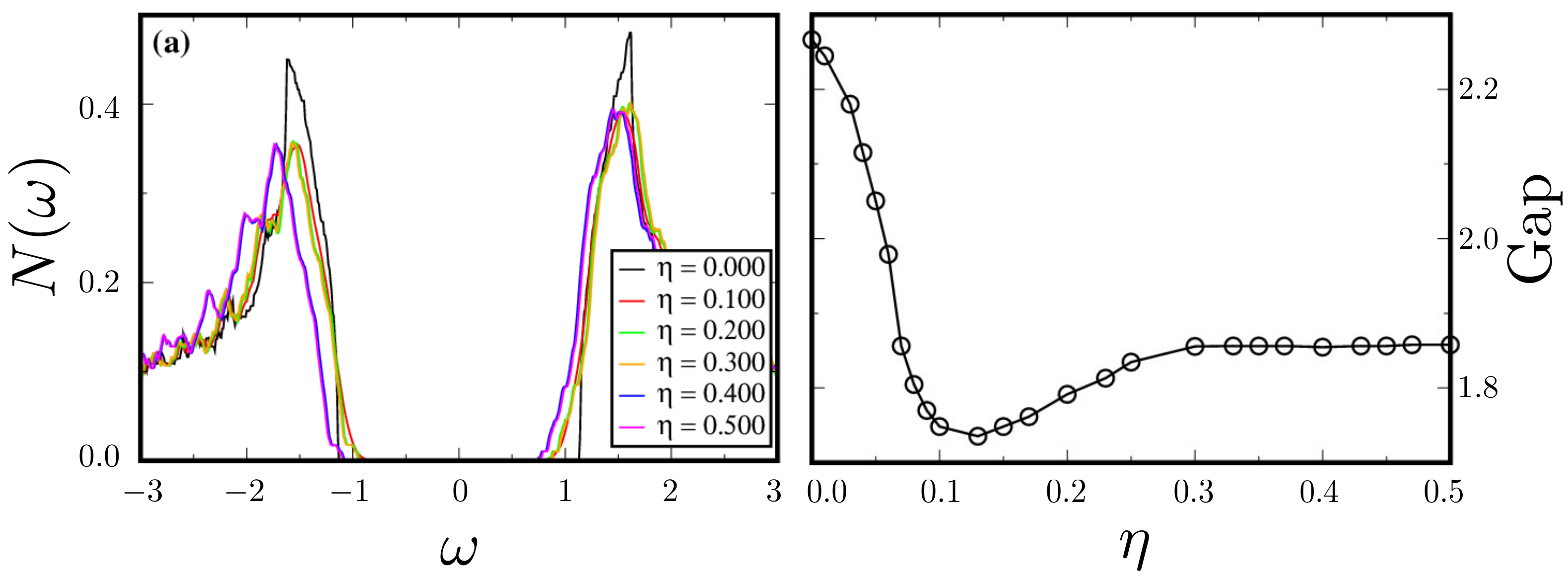}
 \caption{(Left) Electronic density of states vs. impurity concentration with correlated impurities. The data is for $U=4t$, $W=0.1t$ and $t'=0.1t$. (Right) Electronic gap vs. $\eta$. }
\label{fig.gap_eta}
\end{figure}

While SC and CDW compete spatially, their effect on the electronic spectrum is similar. Both SC and CDW order parameters give rise to a full electronic gap at half-filling. We define the electronic density of states as, $N(\omega) = (1/N)\sum_{i}(\mid u_{n}^{i}\mid^2 \delta(\omega-E_{n})+\mid v_{n}^{i}\mid^2
\delta(\omega+E_{n}))$, where, $u_{n}^{i}$ and $v_{n}^{i}$ are the usual BdG eigenvectors \cite{Karmakar2017} and $E_{n}$ are the energy eigenvalues. 
The evolution of the electronic density of states with impurity concentration is shown in Fig.~\ref{fig.gap_eta}. Starting from the clean limit with uniform SC order, impurities introduce gradients in $\Delta$ which reduces the electronic gap. However, after CDW correlations percolate through the system, the CDW order parameter works to increase the gap. The gap, therefore, shows non-monotonic behavior as a function of $\eta$. Similar behavior was found for the gap as a function of magnetic field, due to overlapping vortex cores with CDW order\cite{Karmakar2017}.

\subsection{Loss of pairing}
At large impurity concentrations, $C_\Delta$ vanishes within error bars, indicating the loss of SC order. At this point, this system has strong CDW correlations indicating a genuine CDW phase. In the clean limit, CDW ordering has energy cost which scales as $t'^2$; this has been compensated by the energy gain from impurity potentials here. This region is marked as `CDW' in Fig.~\ref{fig.Cdelta_phi}.

While $C_\Delta$ becomes vanishingly small here, we find that $\Delta$ itself does not uniformly vanish. Rather, SC persists in isolated islands. As these islands are disconnected, the phase of $\Delta$ is no longer uniform through the system. This is reflected in the vanishing of the superfluid stiffness in this regime. 
Within our mean field simulations, we find that both $C_\Delta$ and superfluid stiffness vanish at approximately the same threshold impurity concentration of 
$\eta = 15\%$ for t'=0.1t. 
As $\eta$ is increased further, SC order (albeit in isolated islands) keeps falling as we move closer and closer to a pure CDW state. 





\begin{figure}[t]
\includegraphics[width=\columnwidth]{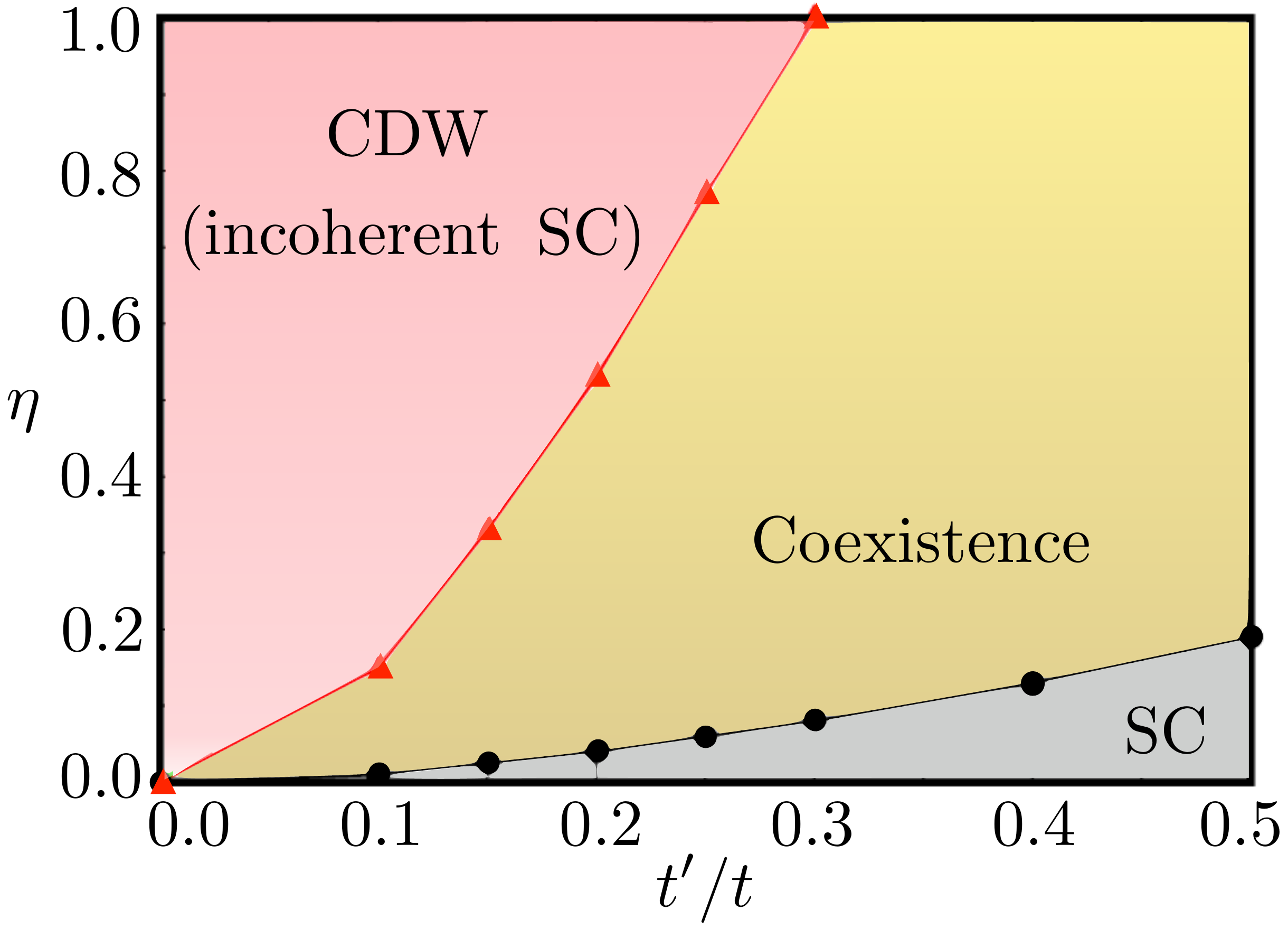}
 \caption{Phase diagram for correlated impurities obtained with $U=4t$ and $W=0.1t$ on a $24\times 24$ lattice. }
\label{fig.eta_tprime_pd}
\end{figure}

\subsection{Phase diagram}
We gather these results into a phase diagram shown in Fig.~\ref{fig.eta_tprime_pd}. We find three phases, as described below. 
At low impurity concentrations, we have a `SC' phase with isolated CDW puddles nucleated by impurities. At a threshold impurity strength, the `coexistence' phase sets in.
The lower boundary of the coexistence region is obtained from the heuristic criterion $C_{\phi}\gtrsim 0.05$. This boundary is well described by $\eta_c(t') \sim t'^2$. This can be rationalized on the basis of the following qualitative argument. Each impurity induces a local CDW texture. This has an associated length scale, $\xi \sim 1/t'$, as shown from field theory arguments above. With many impurities, the average area per impurity is $(L^2 a^2/\eta L^2)=a^2/\eta$, where $\eta$ is the impurity concentration and $a$ is the lattice spacing. The inter-impurity distance scales as $L_{imp-imp}\sim a \eta^{-1/2}$. We expect coexistence to set in when the impurity textures overlap, i.e., $L_{imp-imp} \sim \xi$. This gives us the scaling $\eta_c \sim t'^2$.

\begin{figure*}
\includegraphics[width=3.5in]{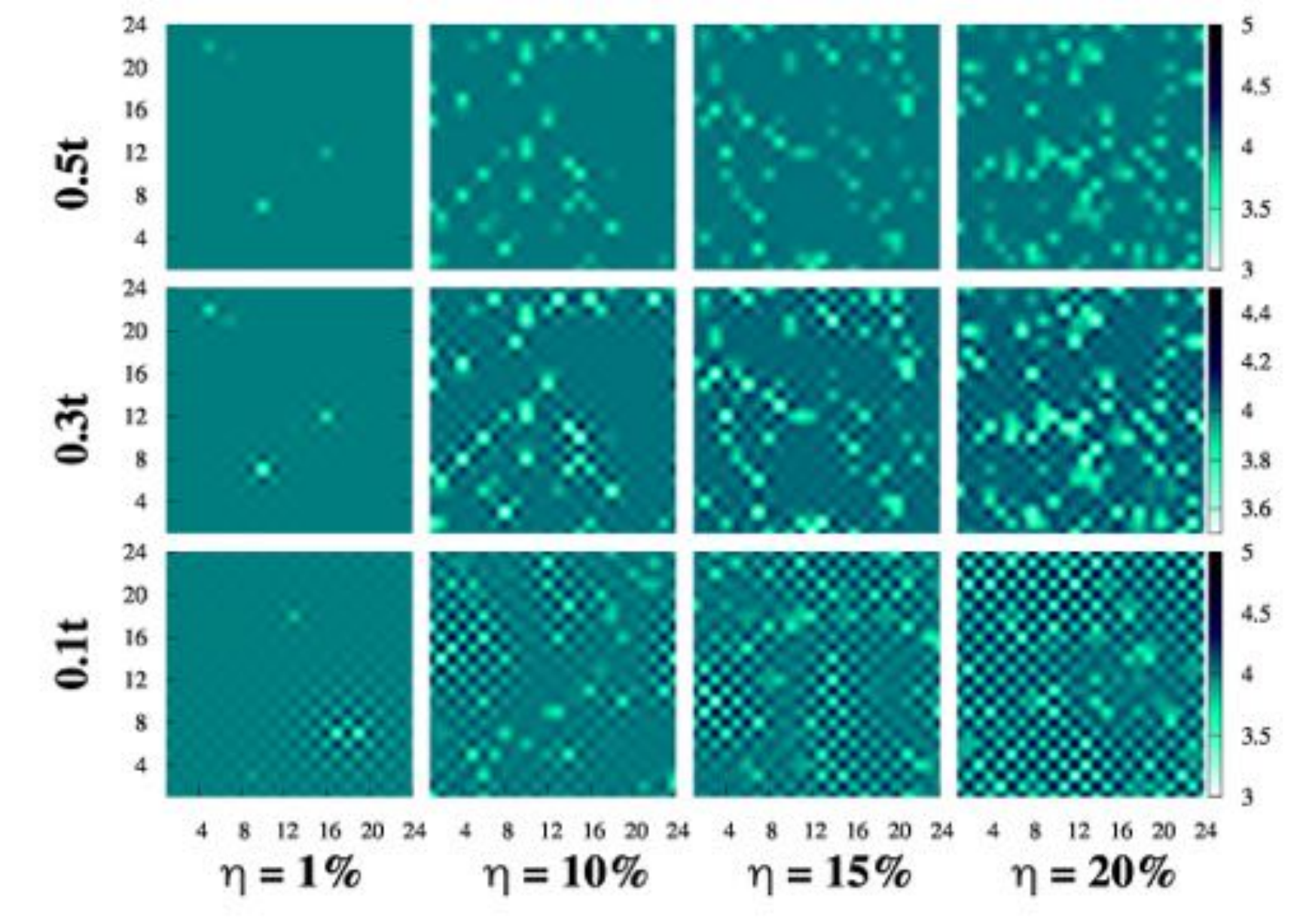}
\includegraphics[width=3.5in]{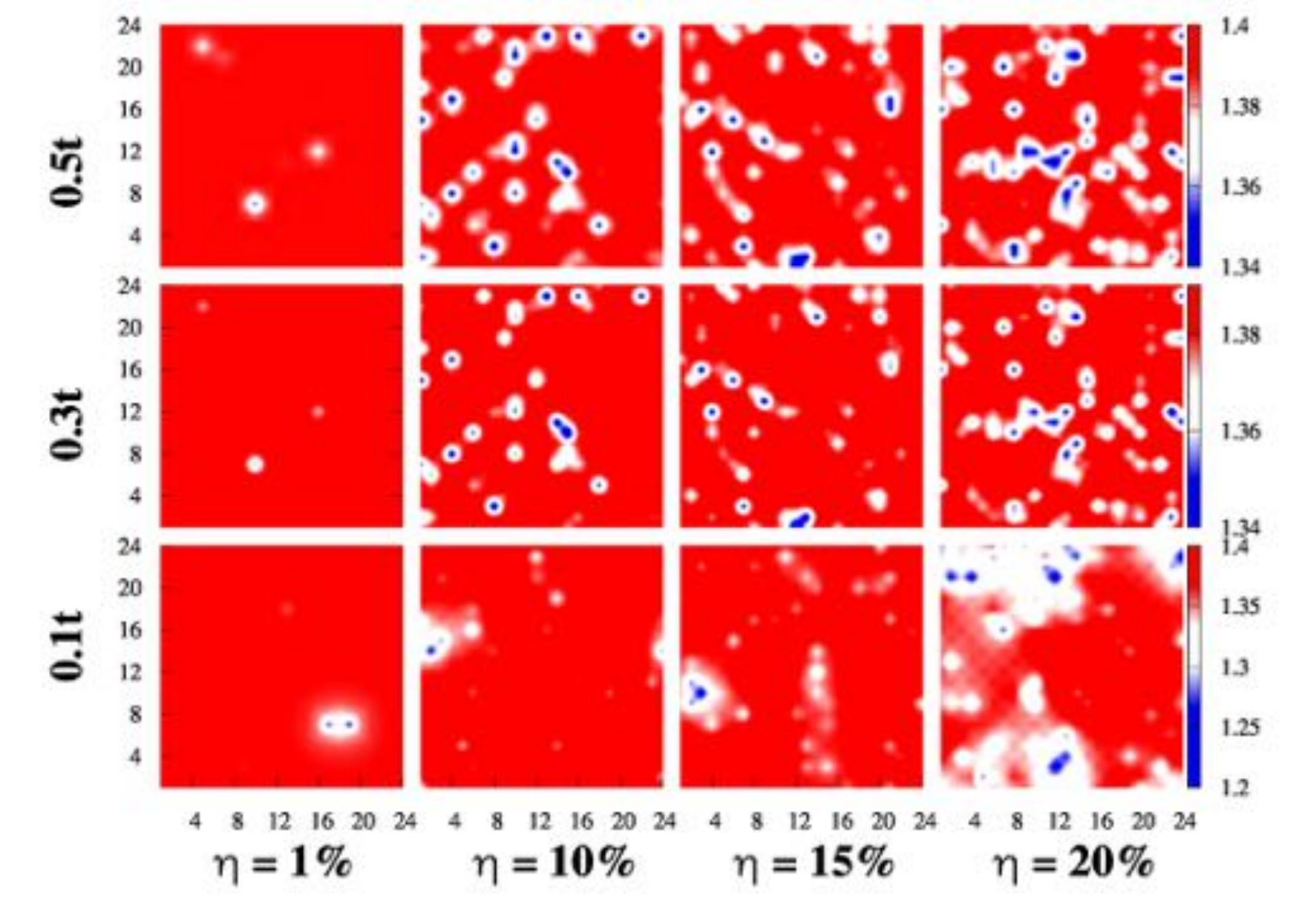}
 \caption{Evolution of order parameters with $t'$ and impurity concentration, $\eta$ with random impurities: (Left) Maps of density. The rows and columns correspond to fixed $t'$ and $\eta$ values respectively. (Right) Maps of $\Delta$, the SC order parameter. 
All panels show data corresponding to $U=4t$ and $W=0.1t$. Each plot shows the mean-field results obtained from a sample disorder configuration. }
\label{fig.random_maps}
\end{figure*}
 
At a large value of $\eta$, coherent SC order is lost. This is signalled by $C_{\Delta}$ vanishing within error bars -- this marks the upper boundary of the `coexistence' phase in the phase diagram of Fig.~\ref{fig.eta_tprime_pd}. However, SC persists beyond this threshold in spatially isolated islands. This can be seen in Fig.~\ref{fig.corr_maps} for $\eta=15\%$ and $t' = 0.1t$. We expect fluctuations beyond mean-field theory to wash out superconductivity in this region. Nevertheless, signatures of pairing may be seen in local probes which are only sensitive to the local pairing amplitude. 

\section{Multiple random impurities}
\label{sec.random}
Here, we discuss random impurity distributions -- we take the Hamiltonian to be 
\begin{eqnarray}
H_{rand.} =H_{t,t',U} + W \sum_{j_r (\eta)}c_{j_r,\sigma}^\dagger c_{j_r,\sigma}. 
\end{eqnarray}
As before, $W>0$ represents the strength of the impurity potential. 
The impurities are randomly positioned within our lattice as we tune the impurity concentration, $\eta$. Here, once again, impurities seed puddles of CDW order. Unlike in the previous section, these CDW puddles may not be in-phase -- they may correspond to different sublattices having higher density. 
Naively, such `random' configurations are the natural setting to study the effect of disorder. 

From our mean-field simulations, we find that random impurity configurations are not conducive to phase coexistence. Typical order parameter configurations for different $t'$ and $\eta$ (impurity concentration) are shown in Fig.~\ref{fig.random_maps}. Clearly, CDW regions do form but do not overlap coherently. This leads to configurations with large order parameter gradients, even for very high impurity concentrations. 

\begin{figure}
\includegraphics[width=\columnwidth]{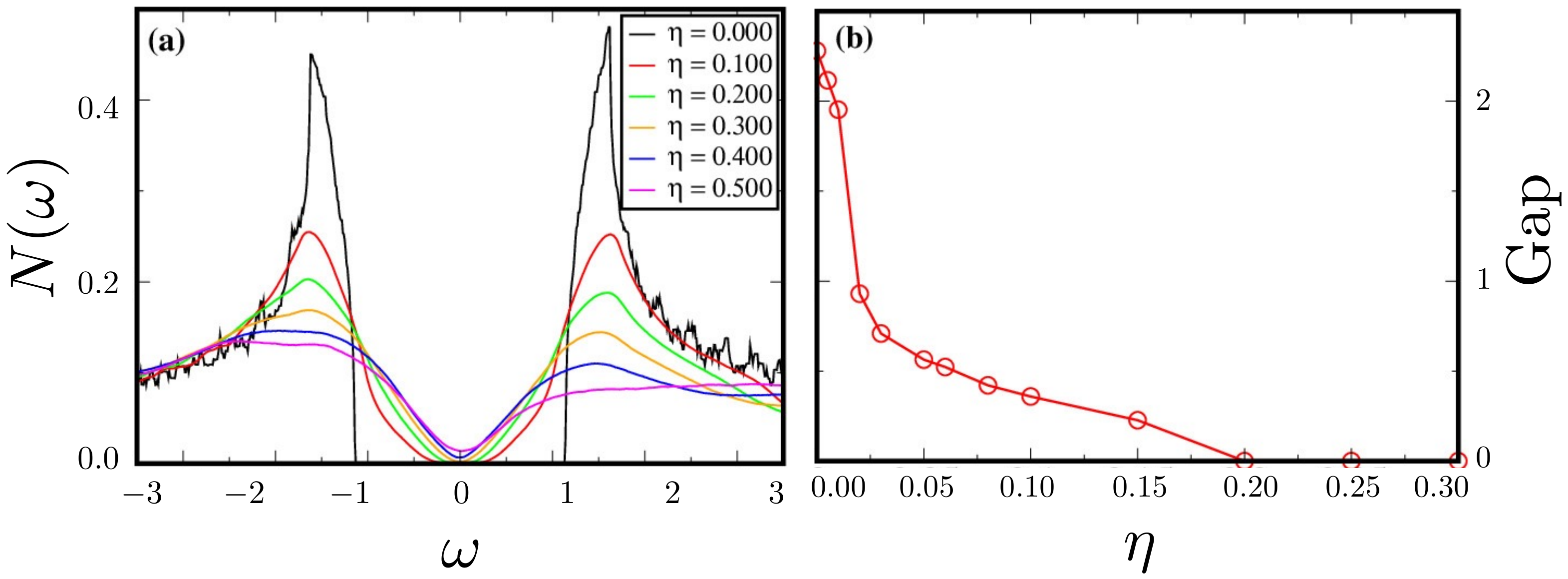}
 \caption{(Left) Electronic density of states vs. impurity concentration with random impurities. The data is for $U=4t$, $W=4.0t$ and $t'=0.1t$. (Right) Electronic gap vs. $\eta$. }
\label{fig.gap_eta_random}
\end{figure}

The lack of coherent ordering is reflected in the electronic spectrum. Fig.~\ref{fig.gap_eta_random} shows the electronic density of states as a function of $\eta$. We see that the gap monotonically decreases with increasing impurity concentration. As the random impurity potential preempts the formation of uniform CDW order, there is no long ranged order that can strengthen the gap. 
For strong impurity potentials with $W\sim U$, at a critical impurity concentration, the gap closes even though the SC order parameter survives in isolated islands. 
This indicates a gapless SC phase which is highly susceptible to pairing fluctuations. Nevertheless, this regime will exhibit order parameter fluctuations that may be visible to local probes.

Our results for the electronic gap may be contrasted with Ref.~\onlinecite{Ghosal2001}, an early Bogoliubov deGennes study of the Hubbard model which found that the fermionic gap does not vanish upon increasing disorder. We first clarify that the impurity scheme used here is very different. Ref.~\onlinecite{Ghosal2001} assumes an impurity potential on every site and tunes the width of the potential distribution. In contrast, we fix the potential strength and tune impurity concentration in order to mimic doped systems. The crucial difference in the physics arises because Ref.~\onlinecite{Ghosal2001} avoids CDW fluctuations by working away from half-filling. In our study, the CDW order provides an extra degree of freedom allowing for stronger order parameter fluctuations. At the threshold where the gap closes, we find that low energy quasiparticle states are localized at regions with strong order parameter gradients, e. g., at the boundary between a strong SC and a strong CDW region.

\section{Impurity correlations driven by ordering}
\label{sec.corr_egy_gain}
In the preceding sections, we considered correlated and random impurity configurations separately. Naively, we may expect impurities in any doped material to be randomly distributed, without any correlations. This intuition is based on a picture of dilute impurities with short range interactions. It is statistically unlikely that impurity centres will be close enough to each other to interact and to thereby develop correlations. Even if the system were to be annealed, it may not lead to significant correlations as impurities may have to travel long distances to feel one another's interaction potential. This is also entropically unlikely.  

However, if the interactions between impurities are long ranged, this argument must be revisited. This is precisely the case in the current problem, with impurities inducing CDW textures with a length scale that can be much greater than the lattice spacing. 
Correlations among impurities allow coherent CDW order to form; as a consequence, a robust electronic gap opens. This leads to a lowering of electronic energy. Moreover, it is easy for impurities to become correlated here. Consider two impurities separated by a distance that is smaller than the coherence length of CDW order. If the impurities are anti-correlated, their CDW puddles will be out-of-phase and incoherent. However, if one of the impurities moves by one lattice spacing, this will suffice to ensure coherent overlap of CDW textures resulting in energy lowering. Thus, in scenarios where impurities seed competing order textures,
we should generically expect correlations between impurities. 

This proposed mechanism for impurity-interactions is directly analogous to RKKY interactions. In the RKKY mechanism\cite{Ruderman1954}, each impurity spin couples to the local spin of conduction electrons. This induces a spin texture. The overlap of these textures lead to an effective impurity-impurity interaction. In our discussion above, an impurity potential can be thought of an impurity-pseudospin which couples to the pseudospin moment of fermions. This results in a pseudospin texture, i.e., a CDW puddle. The overlap of textures lead to an effective interaction between impurities. In the RKKY picture, the conduction electrons involved are low-lying excitations near the Fermi surface. Here, in contrast, the fermions are strongly gapped with no Fermi surface. Nevertheless, there are low energy pseudospin excitations which arise from the proximity to the $SO(3)$ symmetric point. 

\begin{figure}
\includegraphics[width=\columnwidth]{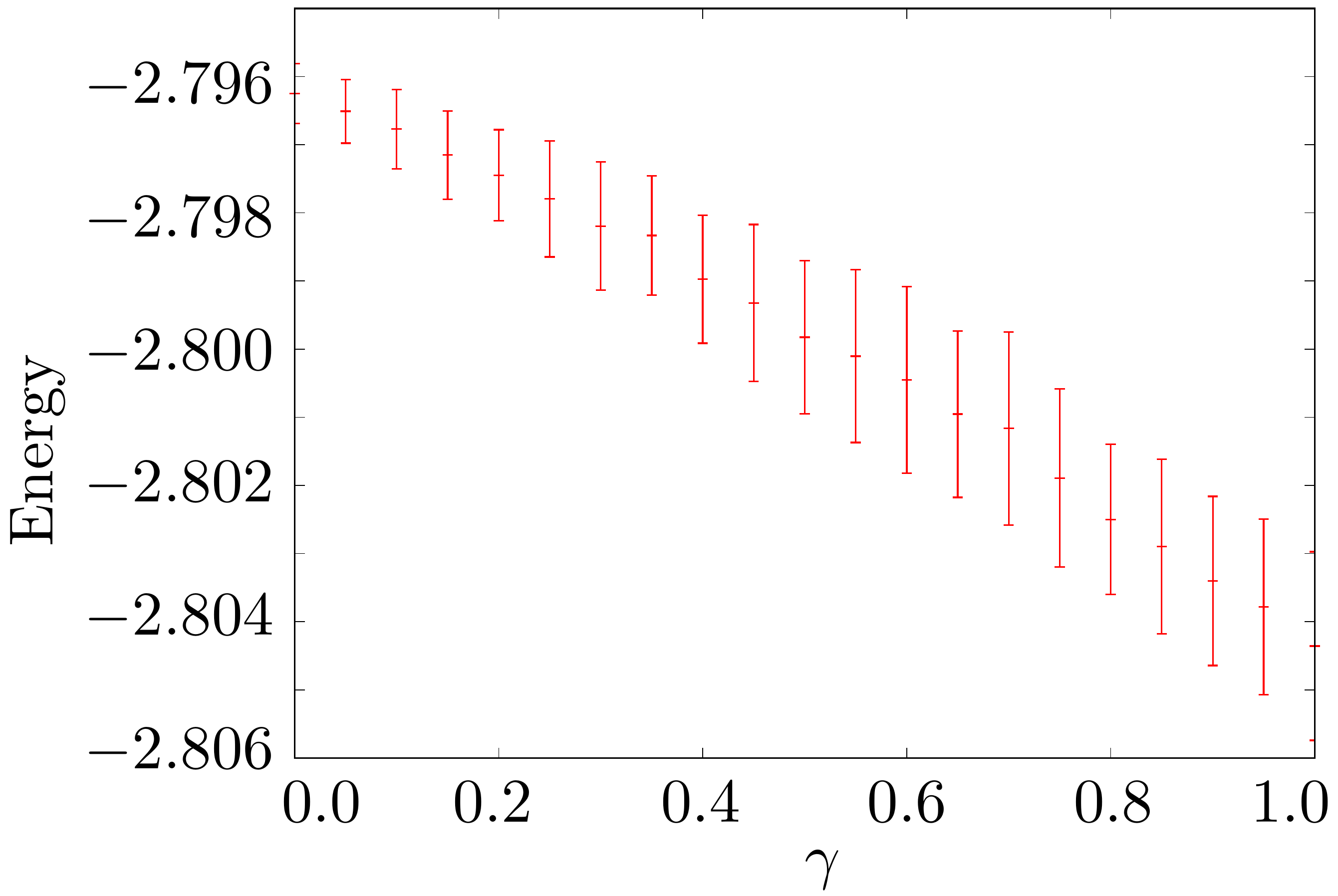}
 \caption{Ground state energy (per site) from mean field theory as a function of $\gamma$, the correlation parameter. At $\gamma = 0$, the impurities are placed randomly. At $\gamma=1$, all impurities are correlated such that all induced CDW textures are in-phase. We show data for a representative case with parameters $U=4t$, $t'=0./1$, $W=0.1$ and $\eta = 0.1$. 
 The error bars are standard deviations after averaging over 100 disorder realizations. }
\label{fig.gamma_egy}
\end{figure}

To quantify this argument, we introduce a parameter $\gamma$ which interpolates between the two scenarios discussed in the previous sections. It represents the fraction of impurities that are placed in a correlated fashion; the remaining impurities are distributed randomly. When $\gamma=0$ and $\gamma=1$, we have random and correlated impurity configurations respectively. For intermediate $\gamma$ values, our impurity configurations smoothly interpolate between these limits. Fig.~\ref{fig.gamma_egy} shows the obtained mean-field ground state energy vs. $\gamma$. As can be seen clearly, energy decreases with correlations between impurities. 

This indicates that the impurities interact with other. Given that a single impurity texture has a length scale $\xi$, we conjecture that the impurity interaction is of the form 
\begin{equation}
H_{imp-imp}\sim -\left\{W_i (-1)^{i} \right\} \left\{ W_j (-1)^{j} \right\} \exp(-\vert \mathbf{r}_i - \mathbf{r}_j\vert/\xi).
\end{equation} 
Here, $W_i$ and $W_j$ are the local potentials. The term $\left\{W_i (-1)^{i} \right\}$ is the coupling to the local CDW field. The length scale $\xi$ depends on the microscopic parameters $t'$ and $U$. In a system with multiple impurities, this interaction will drive the impurities to develop correlations amongst themselves. The time scale for the motion of impurities may be very large -- this will prevent all impurities from becoming correlated. However, we expect that an annealing process will increase the degree of correlation among impurities. 

\section{Summary and discussion}
We have studied the role of impurities in bringing about phase competition in the attractive Hubbard model. To summarize, the Hubbard model with a $t'$ hopping has a SC ground state with low lying CDW excitations. In the strong coupling limit, this is well described by a field theory with two order parameters and a constant-squared-sum constraint. We show that a local impurity is able to induce CDW correlations over a length scale $\xi \sim 1/t'$. 
With many impurities, the induced CDW puddles can overlap giving rise to a `supersolid' phase with coexisting SC and CDW orders. This requires the impurities to be correlated, so that the CDW puddles are in-phase. We argue that the energy lowering from coherent CDW formation will drive such correlations between impurities.  

It is well known that impurities can locally seed competing order\cite{Alloul2009}. We have studied this in a strong coupling limit where order parameters are coupled by a squared-sum constraint. Our results are applicable to any system with a constraint similar to Eq.~\ref{eq.constraint} above, e.g., in the field theories proposed for charge ordered cuprates\cite{Efetov2013,Hayward2014,Wachtel2014}.

The effects of disorder on the Hubbard model have been extensively studied\cite{Ghosal2001,Sakaida2013,Kumar2015,Pasrija2016}. Notably, at the $SO(3)$ symmetric point, random on-site potentials were shown to destroy CDW while sparing SC order\cite{Huscroft1997}. 
A similar finding has been reported for charge order in the cuprates on the basis of a field theory with a squared-sum constraint\cite{Caplan2015}. More general field theory approaches also show that CDW order is destroyed by disorder\cite{Robertson2006,delMaestro2006,Nie2014}.
Our results in Sec.~\ref{sec.random} are in agreement, showing that uncorrelated disorder does not lead to coherent CDW order. However, our study goes beyond this picture and shows that correlations among impurities can develop and stabilize CDW order. The driving mechanism behind this is energy lowering due to coherent electronic ordering. More generally, our study shows that impurities can serve as a tool to manifest competing orders.

Our results resonate with recent experiments on cuprates, especially on YBCO. The dopant Oxygen atoms in YBCO show rich ordering phenomena that is tied to electronic ordering\cite{Bruynseraede1989,Zimmermann2003,Wu2016}. The principle that we have demonstrated in this work -- electronic ordering can have a feedback effect to drive correlations between dopants -- presumably underlies the physics of Oxygen ordering. 
As for the electronic charge ordering itself, there is an ongoing debate as to whether this originates from disorder-induced CDW puddles\cite{Wuincipient2015,Achkar2014}.
Disorder certainly play an important role, e.g., in inducing variations in the local DOS within the long range CDW order obtained in a magnetic field\cite{Zhou2016}. 
Oxygen defects are also believed to pin short range order in Bi-based cuprates\cite{Zeljkovic2012}. Our results suggest that annealed samples will choose an ordering which leads to the lowest electronic energy, i.e., the largest gap at the Fermi surface. 
Also, our work shows that CDW and SC order can coexist in a genuine supersolid. Its properties, collective excitations, etc. are an interesting direction for future studies. 

Our work is also relevant to studies of doped CDW systems, e.g., NbSe$_2$ which shows pseudogap-like behavior\cite{Chatterjee2015}. STM measurements have revealed CDW patches that form around impurities. As the temperature is decreased through $T_{CDW}$, these patches coalesce to form long range order\cite{Arguello2014}.  
Our results suggest that annealing such samples will increase dopant correlations and lead to stronger CDW behavior. There may also be shifts in impurity positions at the onset of long range order so as to allow coherent CDW ordering.
Our results suggest an interesting direction for future experiments: are there macroscopic distinctions between systems with correlated and uncorrelated disorder? For example, can coherent ordering be preempted by increasing randomness in the impurity distribution?

\textit{Acknowledgments:} We thank Arun Paramekanti and David Hawthorn for discussions. We thank HRI, Allahabad for use of the high performance computing cluster facility.

\bibliographystyle{apsrev4-1}
\bibliography{HubbardSO3}

\begin{thebibliography}{37}%
\makeatletter
\providecommand \@ifxundefined [1]{%
 \@ifx{#1\undefined}
}%
\providecommand \@ifnum [1]{%
 \ifnum #1\expandafter \@firstoftwo
 \else \expandafter \@secondoftwo
 \fi
}%
\providecommand \@ifx [1]{%
 \ifx #1\expandafter \@firstoftwo
 \else \expandafter \@secondoftwo
 \fi
}%
\providecommand \natexlab [1]{#1}%
\providecommand \enquote  [1]{``#1''}%
\providecommand \bibnamefont  [1]{#1}%
\providecommand \bibfnamefont [1]{#1}%
\providecommand \citenamefont [1]{#1}%
\providecommand \href@noop [0]{\@secondoftwo}%
\providecommand \href [0]{\begingroup \@sanitize@url \@href}%
\providecommand \@href[1]{\@@startlink{#1}\@@href}%
\providecommand \@@href[1]{\endgroup#1\@@endlink}%
\providecommand \@sanitize@url [0]{\catcode `\\12\catcode `\$12\catcode
  `\&12\catcode `\#12\catcode `\^12\catcode `\_12\catcode `\%12\relax}%
\providecommand \@@startlink[1]{}%
\providecommand \@@endlink[0]{}%
\providecommand \url  [0]{\begingroup\@sanitize@url \@url }%
\providecommand \@url [1]{\endgroup\@href {#1}{\urlprefix }}%
\providecommand \urlprefix  [0]{URL }%
\providecommand \Eprint [0]{\href }%
\providecommand \doibase [0]{http://dx.doi.org/}%
\providecommand \selectlanguage [0]{\@gobble}%
\providecommand \bibinfo  [0]{\@secondoftwo}%
\providecommand \bibfield  [0]{\@secondoftwo}%
\providecommand \translation [1]{[#1]}%
\providecommand \BibitemOpen [0]{}%
\providecommand \bibitemStop [0]{}%
\providecommand \bibitemNoStop [0]{.\EOS\space}%
\providecommand \EOS [0]{\spacefactor3000\relax}%
\providecommand \BibitemShut  [1]{\csname bibitem#1\endcsname}%
\let\auto@bib@innerbib\@empty
\bibitem [{\citenamefont {Lee}\ \emph {et~al.}(2006)\citenamefont {Lee},
  \citenamefont {Nagaosa},\ and\ \citenamefont {Wen}}]{cupratereview}%
  \BibitemOpen
  \bibfield  {author} {\bibinfo {author} {\bibfnamefont {P.~A.}\ \bibnamefont
  {Lee}}, \bibinfo {author} {\bibfnamefont {N.}~\bibnamefont {Nagaosa}}, \ and\
  \bibinfo {author} {\bibfnamefont {X.-G.}\ \bibnamefont {Wen}},\ }\href
  {\doibase 10.1103/RevModPhys.78.17} {\bibfield  {journal} {\bibinfo
  {journal} {Rev. Mod. Phys.}\ }\textbf {\bibinfo {volume} {78}},\ \bibinfo
  {pages} {17} (\bibinfo {year} {2006})}\BibitemShut {NoStop}%
\bibitem [{\citenamefont {{Drew}}\ \emph {et~al.}(2009)\citenamefont {{Drew}},
  \citenamefont {{Niedermayer}}, \citenamefont {{Baker}}, \citenamefont
  {{Pratt}}, \citenamefont {{Blundell}}, \citenamefont {{Lancaster}},
  \citenamefont {{Liu}}, \citenamefont {{Wu}}, \citenamefont {{Chen}},
  \citenamefont {{Watanabe}}, \citenamefont {{Malik}}, \citenamefont
  {{Dubroka}}, \citenamefont {{R{\"o}ssle}}, \citenamefont {{Kim}},
  \citenamefont {{Baines}},\ and\ \citenamefont
  {{Bernhard}}}]{pnictidecoexistence}%
  \BibitemOpen
  \bibfield  {author} {\bibinfo {author} {\bibfnamefont {A.~J.}\ \bibnamefont
  {{Drew}}}, \bibinfo {author} {\bibfnamefont {C.}~\bibnamefont
  {{Niedermayer}}}, \bibinfo {author} {\bibfnamefont {P.~J.}\ \bibnamefont
  {{Baker}}}, \bibinfo {author} {\bibfnamefont {F.~L.}\ \bibnamefont
  {{Pratt}}}, \bibinfo {author} {\bibfnamefont {S.~J.}\ \bibnamefont
  {{Blundell}}}, \bibinfo {author} {\bibfnamefont {T.}~\bibnamefont
  {{Lancaster}}}, \bibinfo {author} {\bibfnamefont {R.~H.}\ \bibnamefont
  {{Liu}}}, \bibinfo {author} {\bibfnamefont {G.}~\bibnamefont {{Wu}}},
  \bibinfo {author} {\bibfnamefont {X.~H.}\ \bibnamefont {{Chen}}}, \bibinfo
  {author} {\bibfnamefont {I.}~\bibnamefont {{Watanabe}}}, \bibinfo {author}
  {\bibfnamefont {V.~K.}\ \bibnamefont {{Malik}}}, \bibinfo {author}
  {\bibfnamefont {A.}~\bibnamefont {{Dubroka}}}, \bibinfo {author}
  {\bibfnamefont {M.}~\bibnamefont {{R{\"o}ssle}}}, \bibinfo {author}
  {\bibfnamefont {K.~W.}\ \bibnamefont {{Kim}}}, \bibinfo {author}
  {\bibfnamefont {C.}~\bibnamefont {{Baines}}}, \ and\ \bibinfo {author}
  {\bibfnamefont {C.}~\bibnamefont {{Bernhard}}},\ }\href {\doibase
  10.1038/nmat2396} {\bibfield  {journal} {\bibinfo  {journal} {Nature
  Materials}\ }\textbf {\bibinfo {volume} {8}},\ \bibinfo {pages} {310}
  (\bibinfo {year} {2009})}\BibitemShut {NoStop}%
\bibitem [{\citenamefont {Morosan}\ \emph {et~al.}(2006)\citenamefont
  {Morosan}, \citenamefont {Zandbergen}, \citenamefont {Dennis}, \citenamefont
  {Bos}, \citenamefont {Onose}, \citenamefont {Klimczuk}, \citenamefont
  {Ramirez}, \citenamefont {Ong},\ and\ \citenamefont {Cava}}]{Morosan2006}%
  \BibitemOpen
  \bibfield  {author} {\bibinfo {author} {\bibfnamefont {E.}~\bibnamefont
  {Morosan}}, \bibinfo {author} {\bibfnamefont {H.~W.}\ \bibnamefont
  {Zandbergen}}, \bibinfo {author} {\bibfnamefont {B.~S.}\ \bibnamefont
  {Dennis}}, \bibinfo {author} {\bibfnamefont {J.~W.~G.}\ \bibnamefont {Bos}},
  \bibinfo {author} {\bibfnamefont {Y.}~\bibnamefont {Onose}}, \bibinfo
  {author} {\bibfnamefont {T.}~\bibnamefont {Klimczuk}}, \bibinfo {author}
  {\bibfnamefont {A.~P.}\ \bibnamefont {Ramirez}}, \bibinfo {author}
  {\bibfnamefont {N.~P.}\ \bibnamefont {Ong}}, \ and\ \bibinfo {author}
  {\bibfnamefont {R.~J.}\ \bibnamefont {Cava}},\ }\href
  {http://dx.doi.org/10.1038/nphys360} {\bibfield  {journal} {\bibinfo
  {journal} {Nat Phys}\ }\textbf {\bibinfo {volume} {2}},\ \bibinfo {pages}
  {544} (\bibinfo {year} {2006})}\BibitemShut {NoStop}%
\bibitem [{\citenamefont {Yang}(1989)}]{Yang1989}%
  \BibitemOpen
  \bibfield  {author} {\bibinfo {author} {\bibfnamefont {C.~N.}\ \bibnamefont
  {Yang}},\ }\href {\doibase 10.1103/PhysRevLett.63.2144} {\bibfield  {journal}
  {\bibinfo  {journal} {Phys. Rev. Lett.}\ }\textbf {\bibinfo {volume} {63}},\
  \bibinfo {pages} {2144} (\bibinfo {year} {1989})}\BibitemShut {NoStop}%
\bibitem [{\citenamefont {Zhang}(1990)}]{Zhang1990}%
  \BibitemOpen
  \bibfield  {author} {\bibinfo {author} {\bibfnamefont {S.}~\bibnamefont
  {Zhang}},\ }\href {\doibase 10.1103/PhysRevLett.65.120} {\bibfield  {journal}
  {\bibinfo  {journal} {Phys. Rev. Lett.}\ }\textbf {\bibinfo {volume} {65}},\
  \bibinfo {pages} {120} (\bibinfo {year} {1990})}\BibitemShut {NoStop}%
\bibitem [{\citenamefont {Yang}\ and\ \citenamefont {Zhang}(1990)}]{Yang1990}%
  \BibitemOpen
  \bibfield  {author} {\bibinfo {author} {\bibfnamefont {C.~N.}\ \bibnamefont
  {Yang}}\ and\ \bibinfo {author} {\bibfnamefont {S.~C.}\ \bibnamefont
  {Zhang}},\ }\href {\doibase 10.1142/S0217984990000933} {\bibfield  {journal}
  {\bibinfo  {journal} {Modern Physics Letters B}\ }\textbf {\bibinfo {volume}
  {04}},\ \bibinfo {pages} {759} (\bibinfo {year} {1990})}\BibitemShut
  {NoStop}%
\bibitem [{\citenamefont {Burkov}\ and\ \citenamefont
  {Paramekanti}(2008)}]{Burkov2008}%
  \BibitemOpen
  \bibfield  {author} {\bibinfo {author} {\bibfnamefont {A.~A.}\ \bibnamefont
  {Burkov}}\ and\ \bibinfo {author} {\bibfnamefont {A.}~\bibnamefont
  {Paramekanti}},\ }\href {\doibase 10.1103/PhysRevLett.100.255301} {\bibfield
  {journal} {\bibinfo  {journal} {Phys. Rev. Lett.}\ }\textbf {\bibinfo
  {volume} {100}},\ \bibinfo {pages} {255301} (\bibinfo {year}
  {2008})}\BibitemShut {NoStop}%
\bibitem [{\citenamefont {Ganesh}\ \emph {et~al.}(2009)\citenamefont {Ganesh},
  \citenamefont {Paramekanti},\ and\ \citenamefont {Burkov}}]{Ganesh2009}%
  \BibitemOpen
  \bibfield  {author} {\bibinfo {author} {\bibfnamefont {R.}~\bibnamefont
  {Ganesh}}, \bibinfo {author} {\bibfnamefont {A.}~\bibnamefont {Paramekanti}},
  \ and\ \bibinfo {author} {\bibfnamefont {A.~A.}\ \bibnamefont {Burkov}},\
  }\href {\doibase 10.1103/PhysRevA.80.043612} {\bibfield  {journal} {\bibinfo
  {journal} {Phys. Rev. A}\ }\textbf {\bibinfo {volume} {80}},\ \bibinfo
  {pages} {043612} (\bibinfo {year} {2009})}\BibitemShut {NoStop}%
\bibitem [{\citenamefont {{Karmakar}}\ \emph {et~al.}(2017)\citenamefont
  {{Karmakar}}, \citenamefont {{Menon}},\ and\ \citenamefont
  {{Ganesh}}}]{Karmakar2017}%
  \BibitemOpen
  \bibfield  {author} {\bibinfo {author} {\bibfnamefont {M.}~\bibnamefont
  {{Karmakar}}}, \bibinfo {author} {\bibfnamefont {G.~I.}\ \bibnamefont
  {{Menon}}}, \ and\ \bibinfo {author} {\bibfnamefont {R.}~\bibnamefont
  {{Ganesh}}},\ }\href@noop {} {\bibfield  {journal} {\bibinfo  {journal}
  {ArXiv e-prints}\ } (\bibinfo {year} {2017})},\ \Eprint
  {http://arxiv.org/abs/1705.01571} {arXiv:1705.01571 [cond-mat.str-el]}
  \BibitemShut {NoStop}%
\bibitem [{\citenamefont {Ramachandran}(2011)}]{Ganeshthesis}%
  \BibitemOpen
  \bibfield  {author} {\bibinfo {author} {\bibfnamefont {G.}~\bibnamefont
  {Ramachandran}},\ }\emph {\bibinfo {title} {Competing Orders in Strongly
  Correlated Systems}},\ \href
  {https://tspace.library.utoronto.ca/handle/1807/32868} {Ph.D. thesis},\
  \bibinfo  {school} {University of Toronto} (\bibinfo {year} {2011}),\
  \bibinfo {note} {chapter 5}\BibitemShut {NoStop}%
\bibitem [{\citenamefont {Arovas}\ \emph {et~al.}(1997)\citenamefont {Arovas},
  \citenamefont {Berlinsky}, \citenamefont {Kallin},\ and\ \citenamefont
  {Zhang}}]{Arovas1997}%
  \BibitemOpen
  \bibfield  {author} {\bibinfo {author} {\bibfnamefont {D.~P.}\ \bibnamefont
  {Arovas}}, \bibinfo {author} {\bibfnamefont {A.~J.}\ \bibnamefont
  {Berlinsky}}, \bibinfo {author} {\bibfnamefont {C.}~\bibnamefont {Kallin}}, \
  and\ \bibinfo {author} {\bibfnamefont {S.-C.}\ \bibnamefont {Zhang}},\ }\href
  {\doibase 10.1103/PhysRevLett.79.2871} {\bibfield  {journal} {\bibinfo
  {journal} {Phys. Rev. Lett.}\ }\textbf {\bibinfo {volume} {79}},\ \bibinfo
  {pages} {2871} (\bibinfo {year} {1997})}\BibitemShut {NoStop}%
\bibitem [{\citenamefont {Demler}\ \emph {et~al.}(2004)\citenamefont {Demler},
  \citenamefont {Hanke},\ and\ \citenamefont {Zhang}}]{Demler2004}%
  \BibitemOpen
  \bibfield  {author} {\bibinfo {author} {\bibfnamefont {E.}~\bibnamefont
  {Demler}}, \bibinfo {author} {\bibfnamefont {W.}~\bibnamefont {Hanke}}, \
  and\ \bibinfo {author} {\bibfnamefont {S.-C.}\ \bibnamefont {Zhang}},\ }\href
  {\doibase 10.1103/RevModPhys.76.909} {\bibfield  {journal} {\bibinfo
  {journal} {Rev. Mod. Phys.}\ }\textbf {\bibinfo {volume} {76}},\ \bibinfo
  {pages} {909} (\bibinfo {year} {2004})}\BibitemShut {NoStop}%
\bibitem [{\citenamefont {Efetov}\ \emph {et~al.}(2013)\citenamefont {Efetov},
  \citenamefont {Meier},\ and\ \citenamefont {Pepin}}]{Efetov2013}%
  \BibitemOpen
  \bibfield  {author} {\bibinfo {author} {\bibfnamefont {K.~B.}\ \bibnamefont
  {Efetov}}, \bibinfo {author} {\bibfnamefont {H.}~\bibnamefont {Meier}}, \
  and\ \bibinfo {author} {\bibfnamefont {C.}~\bibnamefont {Pepin}},\ }\href
  {http://dx.doi.org/10.1038/nphys2641} {\bibfield  {journal} {\bibinfo
  {journal} {Nat Phys}\ }\textbf {\bibinfo {volume} {9}},\ \bibinfo {pages}
  {442} (\bibinfo {year} {2013})}\BibitemShut {NoStop}%
\bibitem [{\citenamefont {Hayward}\ \emph {et~al.}(2014)\citenamefont
  {Hayward}, \citenamefont {Hawthorn}, \citenamefont {Melko},\ and\
  \citenamefont {Sachdev}}]{Hayward2014}%
  \BibitemOpen
  \bibfield  {author} {\bibinfo {author} {\bibfnamefont {L.~E.}\ \bibnamefont
  {Hayward}}, \bibinfo {author} {\bibfnamefont {D.~G.}\ \bibnamefont
  {Hawthorn}}, \bibinfo {author} {\bibfnamefont {R.~G.}\ \bibnamefont {Melko}},
  \ and\ \bibinfo {author} {\bibfnamefont {S.}~\bibnamefont {Sachdev}},\ }\href
  {\doibase 10.1126/science.1246310} {\bibfield  {journal} {\bibinfo  {journal}
  {Science}\ }\textbf {\bibinfo {volume} {343}},\ \bibinfo {pages} {1336}
  (\bibinfo {year} {2014})}\BibitemShut {NoStop}%
\bibitem [{\citenamefont {Wachtel}\ and\ \citenamefont
  {Orgad}(2014)}]{Wachtel2014}%
  \BibitemOpen
  \bibfield  {author} {\bibinfo {author} {\bibfnamefont {G.}~\bibnamefont
  {Wachtel}}\ and\ \bibinfo {author} {\bibfnamefont {D.}~\bibnamefont
  {Orgad}},\ }\href {\doibase 10.1103/PhysRevB.90.224506} {\bibfield  {journal}
  {\bibinfo  {journal} {Phys. Rev. B}\ }\textbf {\bibinfo {volume} {90}},\
  \bibinfo {pages} {224506} (\bibinfo {year} {2014})}\BibitemShut {NoStop}%
\bibitem [{\citenamefont {Boninsegni}\ and\ \citenamefont
  {Prokof'ev}(2012)}]{Boninsegni2012}%
  \BibitemOpen
  \bibfield  {author} {\bibinfo {author} {\bibfnamefont {M.}~\bibnamefont
  {Boninsegni}}\ and\ \bibinfo {author} {\bibfnamefont {N.~V.}\ \bibnamefont
  {Prokof'ev}},\ }\href {\doibase 10.1103/RevModPhys.84.759} {\bibfield
  {journal} {\bibinfo  {journal} {Rev. Mod. Phys.}\ }\textbf {\bibinfo {volume}
  {84}},\ \bibinfo {pages} {759} (\bibinfo {year} {2012})}\BibitemShut
  {NoStop}%
\bibitem [{\citenamefont {Seibold}\ \emph {et~al.}(2012)\citenamefont
  {Seibold}, \citenamefont {Benfatto}, \citenamefont {Castellani},\ and\
  \citenamefont {Lorenzana}}]{castellani2012}%
  \BibitemOpen
  \bibfield  {author} {\bibinfo {author} {\bibfnamefont {G.}~\bibnamefont
  {Seibold}}, \bibinfo {author} {\bibfnamefont {L.}~\bibnamefont {Benfatto}},
  \bibinfo {author} {\bibfnamefont {C.}~\bibnamefont {Castellani}}, \ and\
  \bibinfo {author} {\bibfnamefont {J.}~\bibnamefont {Lorenzana}},\ }\href
  {\doibase 10.1103/PhysRevLett.108.207004} {\bibfield  {journal} {\bibinfo
  {journal} {Phys. Rev. Lett.}\ }\textbf {\bibinfo {volume} {108}},\ \bibinfo
  {pages} {207004} (\bibinfo {year} {2012})}\BibitemShut {NoStop}%
\bibitem [{\citenamefont {Ghosal}\ \emph {et~al.}(2001)\citenamefont {Ghosal},
  \citenamefont {Randeria},\ and\ \citenamefont {Trivedi}}]{Ghosal2001}%
  \BibitemOpen
  \bibfield  {author} {\bibinfo {author} {\bibfnamefont {A.}~\bibnamefont
  {Ghosal}}, \bibinfo {author} {\bibfnamefont {M.}~\bibnamefont {Randeria}}, \
  and\ \bibinfo {author} {\bibfnamefont {N.}~\bibnamefont {Trivedi}},\ }\href
  {\doibase 10.1103/PhysRevB.65.014501} {\bibfield  {journal} {\bibinfo
  {journal} {Phys. Rev. B}\ }\textbf {\bibinfo {volume} {65}},\ \bibinfo
  {pages} {014501} (\bibinfo {year} {2001})}\BibitemShut {NoStop}%
\bibitem [{\citenamefont {Ruderman}\ and\ \citenamefont
  {Kittel}(1954)}]{Ruderman1954}%
  \BibitemOpen
  \bibfield  {author} {\bibinfo {author} {\bibfnamefont {M.~A.}\ \bibnamefont
  {Ruderman}}\ and\ \bibinfo {author} {\bibfnamefont {C.}~\bibnamefont
  {Kittel}},\ }\href {\doibase 10.1103/PhysRev.96.99} {\bibfield  {journal}
  {\bibinfo  {journal} {Phys. Rev.}\ }\textbf {\bibinfo {volume} {96}},\
  \bibinfo {pages} {99} (\bibinfo {year} {1954})}\BibitemShut {NoStop}%
\bibitem [{\citenamefont {Alloul}\ \emph {et~al.}(2009)\citenamefont {Alloul},
  \citenamefont {Bobroff}, \citenamefont {Gabay},\ and\ \citenamefont
  {Hirschfeld}}]{Alloul2009}%
  \BibitemOpen
  \bibfield  {author} {\bibinfo {author} {\bibfnamefont {H.}~\bibnamefont
  {Alloul}}, \bibinfo {author} {\bibfnamefont {J.}~\bibnamefont {Bobroff}},
  \bibinfo {author} {\bibfnamefont {M.}~\bibnamefont {Gabay}}, \ and\ \bibinfo
  {author} {\bibfnamefont {P.~J.}\ \bibnamefont {Hirschfeld}},\ }\href
  {\doibase 10.1103/RevModPhys.81.45} {\bibfield  {journal} {\bibinfo
  {journal} {Rev. Mod. Phys.}\ }\textbf {\bibinfo {volume} {81}},\ \bibinfo
  {pages} {45} (\bibinfo {year} {2009})}\BibitemShut {NoStop}%
\bibitem [{\citenamefont {Sakaida}\ \emph {et~al.}(2013)\citenamefont
  {Sakaida}, \citenamefont {Noda},\ and\ \citenamefont
  {Kawakami}}]{Sakaida2013}%
  \BibitemOpen
  \bibfield  {author} {\bibinfo {author} {\bibfnamefont {M.}~\bibnamefont
  {Sakaida}}, \bibinfo {author} {\bibfnamefont {K.}~\bibnamefont {Noda}}, \
  and\ \bibinfo {author} {\bibfnamefont {N.}~\bibnamefont {Kawakami}},\ }\href
  {\doibase 10.7566/JPSJ.82.074715} {\bibfield  {journal} {\bibinfo  {journal}
  {Journal of the Physical Society of Japan}\ }\textbf {\bibinfo {volume}
  {82}},\ \bibinfo {pages} {074715} (\bibinfo {year} {2013})},\ \Eprint
  {http://arxiv.org/abs/http://dx.doi.org/10.7566/JPSJ.82.074715}
  {http://dx.doi.org/10.7566/JPSJ.82.074715} \BibitemShut {NoStop}%
\bibitem [{\citenamefont {Kumar}\ and\ \citenamefont
  {Chakraborty}(2015)}]{Kumar2015}%
  \BibitemOpen
  \bibfield  {author} {\bibinfo {author} {\bibfnamefont {S.}~\bibnamefont
  {Kumar}}\ and\ \bibinfo {author} {\bibfnamefont {P.~B.}\ \bibnamefont
  {Chakraborty}},\ }\href {\doibase 10.1140/epjb/e2015-50733-2} {\bibfield
  {journal} {\bibinfo  {journal} {The European Physical Journal B}\ }\textbf
  {\bibinfo {volume} {88}},\ \bibinfo {pages} {69} (\bibinfo {year}
  {2015})}\BibitemShut {NoStop}%
\bibitem [{\citenamefont {Pasrija}\ \emph {et~al.}(2016)\citenamefont
  {Pasrija}, \citenamefont {Chakraborty},\ and\ \citenamefont
  {Kumar}}]{Pasrija2016}%
  \BibitemOpen
  \bibfield  {author} {\bibinfo {author} {\bibfnamefont {K.}~\bibnamefont
  {Pasrija}}, \bibinfo {author} {\bibfnamefont {P.~B.}\ \bibnamefont
  {Chakraborty}}, \ and\ \bibinfo {author} {\bibfnamefont {S.}~\bibnamefont
  {Kumar}},\ }\href {\doibase 10.1103/PhysRevB.94.165150} {\bibfield  {journal}
  {\bibinfo  {journal} {Phys. Rev. B}\ }\textbf {\bibinfo {volume} {94}},\
  \bibinfo {pages} {165150} (\bibinfo {year} {2016})}\BibitemShut {NoStop}%
\bibitem [{\citenamefont {Huscroft}\ and\ \citenamefont
  {Scalettar}(1997)}]{Huscroft1997}%
  \BibitemOpen
  \bibfield  {author} {\bibinfo {author} {\bibfnamefont {C.}~\bibnamefont
  {Huscroft}}\ and\ \bibinfo {author} {\bibfnamefont {R.~T.}\ \bibnamefont
  {Scalettar}},\ }\href {\doibase 10.1103/PhysRevB.55.1185} {\bibfield
  {journal} {\bibinfo  {journal} {Phys. Rev. B}\ }\textbf {\bibinfo {volume}
  {55}},\ \bibinfo {pages} {1185} (\bibinfo {year} {1997})}\BibitemShut
  {NoStop}%
\bibitem [{\citenamefont {Caplan}\ \emph {et~al.}(2015)\citenamefont {Caplan},
  \citenamefont {Wachtel},\ and\ \citenamefont {Orgad}}]{Caplan2015}%
  \BibitemOpen
  \bibfield  {author} {\bibinfo {author} {\bibfnamefont {Y.}~\bibnamefont
  {Caplan}}, \bibinfo {author} {\bibfnamefont {G.}~\bibnamefont {Wachtel}}, \
  and\ \bibinfo {author} {\bibfnamefont {D.}~\bibnamefont {Orgad}},\ }\href
  {\doibase 10.1103/PhysRevB.92.224504} {\bibfield  {journal} {\bibinfo
  {journal} {Phys. Rev. B}\ }\textbf {\bibinfo {volume} {92}},\ \bibinfo
  {pages} {224504} (\bibinfo {year} {2015})}\BibitemShut {NoStop}%
\bibitem [{\citenamefont {Robertson}\ \emph {et~al.}(2006)\citenamefont
  {Robertson}, \citenamefont {Kivelson}, \citenamefont {Fradkin}, \citenamefont
  {Fang},\ and\ \citenamefont {Kapitulnik}}]{Robertson2006}%
  \BibitemOpen
  \bibfield  {author} {\bibinfo {author} {\bibfnamefont {J.~A.}\ \bibnamefont
  {Robertson}}, \bibinfo {author} {\bibfnamefont {S.~A.}\ \bibnamefont
  {Kivelson}}, \bibinfo {author} {\bibfnamefont {E.}~\bibnamefont {Fradkin}},
  \bibinfo {author} {\bibfnamefont {A.~C.}\ \bibnamefont {Fang}}, \ and\
  \bibinfo {author} {\bibfnamefont {A.}~\bibnamefont {Kapitulnik}},\ }\href
  {\doibase 10.1103/PhysRevB.74.134507} {\bibfield  {journal} {\bibinfo
  {journal} {Phys. Rev. B}\ }\textbf {\bibinfo {volume} {74}},\ \bibinfo
  {pages} {134507} (\bibinfo {year} {2006})}\BibitemShut {NoStop}%
\bibitem [{\citenamefont {Del~Maestro}\ \emph {et~al.}(2006)\citenamefont
  {Del~Maestro}, \citenamefont {Rosenow},\ and\ \citenamefont
  {Sachdev}}]{delMaestro2006}%
  \BibitemOpen
  \bibfield  {author} {\bibinfo {author} {\bibfnamefont {A.}~\bibnamefont
  {Del~Maestro}}, \bibinfo {author} {\bibfnamefont {B.}~\bibnamefont
  {Rosenow}}, \ and\ \bibinfo {author} {\bibfnamefont {S.}~\bibnamefont
  {Sachdev}},\ }\href {\doibase 10.1103/PhysRevB.74.024520} {\bibfield
  {journal} {\bibinfo  {journal} {Phys. Rev. B}\ }\textbf {\bibinfo {volume}
  {74}},\ \bibinfo {pages} {024520} (\bibinfo {year} {2006})}\BibitemShut
  {NoStop}%
\bibitem [{\citenamefont {Nie}\ \emph {et~al.}(2014)\citenamefont {Nie},
  \citenamefont {Tarjus},\ and\ \citenamefont {Kivelson}}]{Nie2014}%
  \BibitemOpen
  \bibfield  {author} {\bibinfo {author} {\bibfnamefont {L.}~\bibnamefont
  {Nie}}, \bibinfo {author} {\bibfnamefont {G.}~\bibnamefont {Tarjus}}, \ and\
  \bibinfo {author} {\bibfnamefont {S.~A.}\ \bibnamefont {Kivelson}},\ }\href
  {\doibase 10.1073/pnas.1406019111} {\bibfield  {journal} {\bibinfo  {journal}
  {Proceedings of the National Academy of Sciences}\ }\textbf {\bibinfo
  {volume} {111}},\ \bibinfo {pages} {7980} (\bibinfo {year} {2014})},\ \Eprint
  {http://arxiv.org/abs/http://www.pnas.org/content/111/22/7980.full.pdf}
  {http://www.pnas.org/content/111/22/7980.full.pdf} \BibitemShut {NoStop}%
\bibitem [{\citenamefont {Bruynseraede}\ \emph {et~al.}(1989)\citenamefont
  {Bruynseraede}, \citenamefont {Vanacken}, \citenamefont {Wuyts},
  \citenamefont {van Haesendonck}, \citenamefont {Locquet},\ and\ \citenamefont
  {Schuller}}]{Bruynseraede1989}%
  \BibitemOpen
  \bibfield  {author} {\bibinfo {author} {\bibfnamefont {Y.}~\bibnamefont
  {Bruynseraede}}, \bibinfo {author} {\bibfnamefont {J.}~\bibnamefont
  {Vanacken}}, \bibinfo {author} {\bibfnamefont {B.}~\bibnamefont {Wuyts}},
  \bibinfo {author} {\bibfnamefont {C.}~\bibnamefont {van Haesendonck}},
  \bibinfo {author} {\bibfnamefont {J.-P.}\ \bibnamefont {Locquet}}, \ and\
  \bibinfo {author} {\bibfnamefont {I.~K.}\ \bibnamefont {Schuller}},\ }\href
  {http://stacks.iop.org/1402-4896/1989/i=T29/a=018} {\bibfield  {journal}
  {\bibinfo  {journal} {Physica Scripta}\ }\textbf {\bibinfo {volume} {1989}},\
  \bibinfo {pages} {100} (\bibinfo {year} {1989})}\BibitemShut {NoStop}%
\bibitem [{\citenamefont {Zimmermann}\ \emph {et~al.}(2003)\citenamefont
  {Zimmermann}, \citenamefont {Schneider}, \citenamefont {Frello},
  \citenamefont {Andersen}, \citenamefont {Madsen}, \citenamefont {K\"all},
  \citenamefont {Poulsen}, \citenamefont {Liang}, \citenamefont {Dosanjh},\
  and\ \citenamefont {Hardy}}]{Zimmermann2003}%
  \BibitemOpen
  \bibfield  {author} {\bibinfo {author} {\bibfnamefont {M.~v.}\ \bibnamefont
  {Zimmermann}}, \bibinfo {author} {\bibfnamefont {J.~R.}\ \bibnamefont
  {Schneider}}, \bibinfo {author} {\bibfnamefont {T.}~\bibnamefont {Frello}},
  \bibinfo {author} {\bibfnamefont {N.~H.}\ \bibnamefont {Andersen}}, \bibinfo
  {author} {\bibfnamefont {J.}~\bibnamefont {Madsen}}, \bibinfo {author}
  {\bibfnamefont {M.}~\bibnamefont {K\"all}}, \bibinfo {author} {\bibfnamefont
  {H.~F.}\ \bibnamefont {Poulsen}}, \bibinfo {author} {\bibfnamefont
  {R.}~\bibnamefont {Liang}}, \bibinfo {author} {\bibfnamefont
  {P.}~\bibnamefont {Dosanjh}}, \ and\ \bibinfo {author} {\bibfnamefont
  {W.~N.}\ \bibnamefont {Hardy}},\ }\href {\doibase 10.1103/PhysRevB.68.104515}
  {\bibfield  {journal} {\bibinfo  {journal} {Phys. Rev. B}\ }\textbf {\bibinfo
  {volume} {68}},\ \bibinfo {pages} {104515} (\bibinfo {year}
  {2003})}\BibitemShut {NoStop}%
\bibitem [{\citenamefont {Wu}\ \emph {et~al.}(2016)\citenamefont {Wu},
  \citenamefont {Zhou}, \citenamefont {Hirata}, \citenamefont {Vinograd},
  \citenamefont {Mayaffre}, \citenamefont {Liang}, \citenamefont {Hardy},
  \citenamefont {Bonn}, \citenamefont {Loew}, \citenamefont {Porras},
  \citenamefont {Haug}, \citenamefont {Lin}, \citenamefont {Hinkov},
  \citenamefont {Keimer},\ and\ \citenamefont {Julien}}]{Wu2016}%
  \BibitemOpen
  \bibfield  {author} {\bibinfo {author} {\bibfnamefont {T.}~\bibnamefont
  {Wu}}, \bibinfo {author} {\bibfnamefont {R.}~\bibnamefont {Zhou}}, \bibinfo
  {author} {\bibfnamefont {M.}~\bibnamefont {Hirata}}, \bibinfo {author}
  {\bibfnamefont {I.}~\bibnamefont {Vinograd}}, \bibinfo {author}
  {\bibfnamefont {H.}~\bibnamefont {Mayaffre}}, \bibinfo {author}
  {\bibfnamefont {R.}~\bibnamefont {Liang}}, \bibinfo {author} {\bibfnamefont
  {W.~N.}\ \bibnamefont {Hardy}}, \bibinfo {author} {\bibfnamefont {D.~A.}\
  \bibnamefont {Bonn}}, \bibinfo {author} {\bibfnamefont {T.}~\bibnamefont
  {Loew}}, \bibinfo {author} {\bibfnamefont {J.}~\bibnamefont {Porras}},
  \bibinfo {author} {\bibfnamefont {D.}~\bibnamefont {Haug}}, \bibinfo {author}
  {\bibfnamefont {C.~T.}\ \bibnamefont {Lin}}, \bibinfo {author} {\bibfnamefont
  {V.}~\bibnamefont {Hinkov}}, \bibinfo {author} {\bibfnamefont
  {B.}~\bibnamefont {Keimer}}, \ and\ \bibinfo {author} {\bibfnamefont {M.-H.}\
  \bibnamefont {Julien}},\ }\href {\doibase 10.1103/PhysRevB.93.134518}
  {\bibfield  {journal} {\bibinfo  {journal} {Phys. Rev. B}\ }\textbf {\bibinfo
  {volume} {93}},\ \bibinfo {pages} {134518} (\bibinfo {year}
  {2016})}\BibitemShut {NoStop}%
\bibitem [{\citenamefont {{Wu}}\ \emph {et~al.}(2015)\citenamefont {{Wu}},
  \citenamefont {{Mayaffre}}, \citenamefont {{Kr{\"a}mer}}, \citenamefont
  {{Horvati{\'c}}}, \citenamefont {{Berthier}}, \citenamefont {{Hardy}},
  \citenamefont {{Liang}}, \citenamefont {{Bonn}},\ and\ \citenamefont
  {{Julien}}}]{Wuincipient2015}%
  \BibitemOpen
  \bibfield  {author} {\bibinfo {author} {\bibfnamefont {T.}~\bibnamefont
  {{Wu}}}, \bibinfo {author} {\bibfnamefont {H.}~\bibnamefont {{Mayaffre}}},
  \bibinfo {author} {\bibfnamefont {S.}~\bibnamefont {{Kr{\"a}mer}}}, \bibinfo
  {author} {\bibfnamefont {M.}~\bibnamefont {{Horvati{\'c}}}}, \bibinfo
  {author} {\bibfnamefont {C.}~\bibnamefont {{Berthier}}}, \bibinfo {author}
  {\bibfnamefont {W.~N.}\ \bibnamefont {{Hardy}}}, \bibinfo {author}
  {\bibfnamefont {R.}~\bibnamefont {{Liang}}}, \bibinfo {author} {\bibfnamefont
  {D.~A.}\ \bibnamefont {{Bonn}}}, \ and\ \bibinfo {author} {\bibfnamefont
  {M.-H.}\ \bibnamefont {{Julien}}},\ }\href {\doibase 10.1038/ncomms7438}
  {\bibfield  {journal} {\bibinfo  {journal} {Nature Communications}\ }\textbf
  {\bibinfo {volume} {6}},\ \bibinfo {eid} {6438} (\bibinfo {year} {2015})},\
  \Eprint {http://arxiv.org/abs/1404.1617} {arXiv:1404.1617
  [cond-mat.supr-con]} \BibitemShut {NoStop}%
\bibitem [{\citenamefont {Achkar}\ \emph {et~al.}(2014)\citenamefont {Achkar},
  \citenamefont {Mao}, \citenamefont {McMahon}, \citenamefont {Sutarto},
  \citenamefont {He}, \citenamefont {Liang}, \citenamefont {Bonn},
  \citenamefont {Hardy},\ and\ \citenamefont {Hawthorn}}]{Achkar2014}%
  \BibitemOpen
  \bibfield  {author} {\bibinfo {author} {\bibfnamefont {A.~J.}\ \bibnamefont
  {Achkar}}, \bibinfo {author} {\bibfnamefont {X.}~\bibnamefont {Mao}},
  \bibinfo {author} {\bibfnamefont {C.}~\bibnamefont {McMahon}}, \bibinfo
  {author} {\bibfnamefont {R.}~\bibnamefont {Sutarto}}, \bibinfo {author}
  {\bibfnamefont {F.}~\bibnamefont {He}}, \bibinfo {author} {\bibfnamefont
  {R.}~\bibnamefont {Liang}}, \bibinfo {author} {\bibfnamefont {D.~A.}\
  \bibnamefont {Bonn}}, \bibinfo {author} {\bibfnamefont {W.~N.}\ \bibnamefont
  {Hardy}}, \ and\ \bibinfo {author} {\bibfnamefont {D.~G.}\ \bibnamefont
  {Hawthorn}},\ }\href {\doibase 10.1103/PhysRevLett.113.107002} {\bibfield
  {journal} {\bibinfo  {journal} {Phys. Rev. Lett.}\ }\textbf {\bibinfo
  {volume} {113}},\ \bibinfo {pages} {107002} (\bibinfo {year}
  {2014})}\BibitemShut {NoStop}%
\bibitem [{\citenamefont {Zhou}\ \emph {et~al.}(2017)\citenamefont {Zhou},
  \citenamefont {Hirata}, \citenamefont {Wu}, \citenamefont {Vinograd},
  \citenamefont {Mayaffre}, \citenamefont {Kr\"amer}, \citenamefont
  {Horvati\ifmmode~\acute{c}\else \'{c}\fi{}}, \citenamefont {Berthier},
  \citenamefont {Reyes}, \citenamefont {Kuhns}, \citenamefont {Liang},
  \citenamefont {Hardy}, \citenamefont {Bonn},\ and\ \citenamefont
  {Julien}}]{Zhou2016}%
  \BibitemOpen
  \bibfield  {author} {\bibinfo {author} {\bibfnamefont {R.}~\bibnamefont
  {Zhou}}, \bibinfo {author} {\bibfnamefont {M.}~\bibnamefont {Hirata}},
  \bibinfo {author} {\bibfnamefont {T.}~\bibnamefont {Wu}}, \bibinfo {author}
  {\bibfnamefont {I.}~\bibnamefont {Vinograd}}, \bibinfo {author}
  {\bibfnamefont {H.}~\bibnamefont {Mayaffre}}, \bibinfo {author}
  {\bibfnamefont {S.}~\bibnamefont {Kr\"amer}}, \bibinfo {author}
  {\bibfnamefont {M.}~\bibnamefont {Horvati\ifmmode~\acute{c}\else
  \'{c}\fi{}}}, \bibinfo {author} {\bibfnamefont {C.}~\bibnamefont {Berthier}},
  \bibinfo {author} {\bibfnamefont {A.~P.}\ \bibnamefont {Reyes}}, \bibinfo
  {author} {\bibfnamefont {P.~L.}\ \bibnamefont {Kuhns}}, \bibinfo {author}
  {\bibfnamefont {R.}~\bibnamefont {Liang}}, \bibinfo {author} {\bibfnamefont
  {W.~N.}\ \bibnamefont {Hardy}}, \bibinfo {author} {\bibfnamefont {D.~A.}\
  \bibnamefont {Bonn}}, \ and\ \bibinfo {author} {\bibfnamefont {M.-H.}\
  \bibnamefont {Julien}},\ }\href {\doibase 10.1103/PhysRevLett.118.017001}
  {\bibfield  {journal} {\bibinfo  {journal} {Phys. Rev. Lett.}\ }\textbf
  {\bibinfo {volume} {118}},\ \bibinfo {pages} {017001} (\bibinfo {year}
  {2017})}\BibitemShut {NoStop}%
\bibitem [{\citenamefont {Zeljkovic}\ \emph {et~al.}(2012)\citenamefont
  {Zeljkovic}, \citenamefont {Xu}, \citenamefont {Wen}, \citenamefont {Gu},
  \citenamefont {Markiewicz},\ and\ \citenamefont {Hoffman}}]{Zeljkovic2012}%
  \BibitemOpen
  \bibfield  {author} {\bibinfo {author} {\bibfnamefont {I.}~\bibnamefont
  {Zeljkovic}}, \bibinfo {author} {\bibfnamefont {Z.}~\bibnamefont {Xu}},
  \bibinfo {author} {\bibfnamefont {J.}~\bibnamefont {Wen}}, \bibinfo {author}
  {\bibfnamefont {G.}~\bibnamefont {Gu}}, \bibinfo {author} {\bibfnamefont
  {R.~S.}\ \bibnamefont {Markiewicz}}, \ and\ \bibinfo {author} {\bibfnamefont
  {J.~E.}\ \bibnamefont {Hoffman}},\ }\href {\doibase 10.1126/science.1218648}
  {\bibfield  {journal} {\bibinfo  {journal} {Science}\ }\textbf {\bibinfo
  {volume} {337}},\ \bibinfo {pages} {320} (\bibinfo {year} {2012})},\ \Eprint
  {http://arxiv.org/abs/http://science.sciencemag.org/content/337/6092/320.full.pdf}
  {http://science.sciencemag.org/content/337/6092/320.full.pdf} \BibitemShut
  {NoStop}%
\bibitem [{\citenamefont {Chatterjee}\ \emph {et~al.}(2015)\citenamefont
  {Chatterjee}, \citenamefont {Zhao}, \citenamefont {Iavarone}, \citenamefont
  {Di~Capua}, \citenamefont {Castellan}, \citenamefont {Karapetrov},
  \citenamefont {Malliakas}, \citenamefont {Kanatzidis}, \citenamefont {Claus},
  \citenamefont {Ruff}, \citenamefont {Weber}, \citenamefont {van Wezel},
  \citenamefont {Campuzano}, \citenamefont {Osborn}, \citenamefont {Randeria},
  \citenamefont {Trivedi}, \citenamefont {Norman},\ and\ \citenamefont
  {Rosenkranz}}]{Chatterjee2015}%
  \BibitemOpen
  \bibfield  {author} {\bibinfo {author} {\bibfnamefont {U.}~\bibnamefont
  {Chatterjee}}, \bibinfo {author} {\bibfnamefont {J.}~\bibnamefont {Zhao}},
  \bibinfo {author} {\bibfnamefont {M.}~\bibnamefont {Iavarone}}, \bibinfo
  {author} {\bibfnamefont {R.}~\bibnamefont {Di~Capua}}, \bibinfo {author}
  {\bibfnamefont {J.~P.}\ \bibnamefont {Castellan}}, \bibinfo {author}
  {\bibfnamefont {G.}~\bibnamefont {Karapetrov}}, \bibinfo {author}
  {\bibfnamefont {C.~D.}\ \bibnamefont {Malliakas}}, \bibinfo {author}
  {\bibfnamefont {M.~G.}\ \bibnamefont {Kanatzidis}}, \bibinfo {author}
  {\bibfnamefont {H.}~\bibnamefont {Claus}}, \bibinfo {author} {\bibfnamefont
  {J.~P.~C.}\ \bibnamefont {Ruff}}, \bibinfo {author} {\bibfnamefont
  {F.}~\bibnamefont {Weber}}, \bibinfo {author} {\bibfnamefont
  {J.}~\bibnamefont {van Wezel}}, \bibinfo {author} {\bibfnamefont {J.~C.}\
  \bibnamefont {Campuzano}}, \bibinfo {author} {\bibfnamefont {R.}~\bibnamefont
  {Osborn}}, \bibinfo {author} {\bibfnamefont {M.}~\bibnamefont {Randeria}},
  \bibinfo {author} {\bibfnamefont {N.}~\bibnamefont {Trivedi}}, \bibinfo
  {author} {\bibfnamefont {M.~R.}\ \bibnamefont {Norman}}, \ and\ \bibinfo
  {author} {\bibfnamefont {S.}~\bibnamefont {Rosenkranz}},\ }\href
  {http://dx.doi.org/10.1038/ncomms7313} {\ \textbf {\bibinfo {volume} {6}},\
  \bibinfo {pages} {6313 EP } (\bibinfo {year} {2015})},\ \bibinfo {note}
  {article}\BibitemShut {NoStop}%
\bibitem [{\citenamefont {Arguello}\ \emph {et~al.}(2014)\citenamefont
  {Arguello}, \citenamefont {Chockalingam}, \citenamefont {Rosenthal},
  \citenamefont {Zhao}, \citenamefont {Guti\'errez}, \citenamefont {Kang},
  \citenamefont {Chung}, \citenamefont {Fernandes}, \citenamefont {Jia},
  \citenamefont {Millis}, \citenamefont {Cava},\ and\ \citenamefont
  {Pasupathy}}]{Arguello2014}%
  \BibitemOpen
  \bibfield  {author} {\bibinfo {author} {\bibfnamefont {C.~J.}\ \bibnamefont
  {Arguello}}, \bibinfo {author} {\bibfnamefont {S.~P.}\ \bibnamefont
  {Chockalingam}}, \bibinfo {author} {\bibfnamefont {E.~P.}\ \bibnamefont
  {Rosenthal}}, \bibinfo {author} {\bibfnamefont {L.}~\bibnamefont {Zhao}},
  \bibinfo {author} {\bibfnamefont {C.}~\bibnamefont {Guti\'errez}}, \bibinfo
  {author} {\bibfnamefont {J.~H.}\ \bibnamefont {Kang}}, \bibinfo {author}
  {\bibfnamefont {W.~C.}\ \bibnamefont {Chung}}, \bibinfo {author}
  {\bibfnamefont {R.~M.}\ \bibnamefont {Fernandes}}, \bibinfo {author}
  {\bibfnamefont {S.}~\bibnamefont {Jia}}, \bibinfo {author} {\bibfnamefont
  {A.~J.}\ \bibnamefont {Millis}}, \bibinfo {author} {\bibfnamefont {R.~J.}\
  \bibnamefont {Cava}}, \ and\ \bibinfo {author} {\bibfnamefont {A.~N.}\
  \bibnamefont {Pasupathy}},\ }\href {\doibase 10.1103/PhysRevB.89.235115}
  {\bibfield  {journal} {\bibinfo  {journal} {Phys. Rev. B}\ }\textbf {\bibinfo
  {volume} {89}},\ \bibinfo {pages} {235115} (\bibinfo {year}
  {2014})}\BibitemShut {NoStop}%
\end{thebibliography}%

\end{document}